\documentstyle[preprint,tighten,eqsecnum,aps]{revtex}
\begin{document}
\draft

\hyphenation{
mini-su-per-space
pre-factor
mani-fold
mani-folds
holo-mor-phic
anti-holo-mor-phic
}

\def\cH{{\cal{H}}}
\def\cHh{{\hat{\cal{H}}}}
\def\cD{{\cal{D}}}
\def\half{{1\over2}}
\def\casehalf{{\case{1}{2}}}

\def\zi{{z_I}}
\def\zbi{{{\bar z}_I}}
\def\zj{{z_J}}
\def\zbj{{{\bar{z}}_J}}
\def\zhi{{{\hat{z}}_I}}
\def\zbhi{{{\hat{\bar{z}}}_I}}
\def\zhj{{{\hat{z}}_J}}
\def\zbhj{{{\hat{\bar{z}}}_J}}

\def\vaholo{{V_{\rm AA}}}

\def\vholo{{V_{\rm HH}}}
\def\vholophy{{V_{\rm HH,0}}}
\def\vtholophy{{{\tilde V}_{\rm HH,0}}}
\def\vpholo{{V^+_{\rm HH}}}
\def\vmholo{{V^-_{\rm HH}}}
\def\vpmholo{{V^\pm_{\rm HH}}}
\def\vepsholo{{V^\epsilon_{\rm HH}}}

\def\hilpholo{{{\cal{H}}^+_{\rm HH}}}
\def\hilmholo{{{\cal{H}}^-_{\rm HH}}}
\def\hilpmholo{{{\cal{H}}^\pm_{\rm HH}}}
\def\hilepsholo{{{\cal{H}}^\epsilon_{\rm HH}}}

\def\vmix{{V_{\rm HA}}}
\def\vmixphy{{V_{\rm HA,0}}}
\def\vtmixphy{{{\tilde V}_{\rm HA,0}}}
\def\vonemix{{V^1_{\rm HA}}}
\def\vtwomix{{V^2_{\rm HA}}}
\def\vepsmix{{V^\epsilon_{\rm HA}}}

\def\hilonemix{{{\cal{H}}^1_{\rm HA}}}
\def\hiltwomix{{{\cal{H}}^2_{\rm HA}}}
\def\hilepsmix{{{\cal{H}}^\epsilon_{\rm HA}}}

\def\vpos{{V_{\rm pos}}}
\def\vtposphy{{{\tilde V}_{\rm pos,0}}}
\def\vonepos{{V^{A,B}_{\rm pos}}}

\def\astarphy{{{\cal{A}}^{({\star})}_{\rm phy}}}
\def\tastarphy{{{\tilde{\cal{A}}}^{({\star})}_{\rm phy}}}

\def\jhpm{{{\hat{J}}_{\pm}}}
\def\jhp{{{\hat{J}}_{+}}}
\def\jhm{{{\hat{J}}_{-}}}
\def\jho{{{\hat{J}}_{0}}}

\def\lhpm{{{\hat{L}}_{\pm}}}
\def\lhp{{{\hat{L}}_{+}}}
\def\lhm{{{\hat{L}}_{-}}}
\def\lho{{{\hat{L}}_{0}}}

\def\jhx{{{\hat{J}}_{x}}}
\def\jhy{{{\hat{J}}_{y}}}

\def\phibetas{{\Phi^\pm_{{\bar \beta}_I}}}
\def\phibetaplus{{\Phi^+_{{\bar \beta}_I}}}
\def\phibetaminus{{\Phi^-_{{\bar \beta}_I}}}
\def\phialphas{{\Phi^\pm_{{\bar \alpha}_I}}}

\def\psibetas{{\Psi_{{\bar\beta}_1,\beta_2}}}
\def\psialphas{{\Psi_{{\bar\alpha}_1,\alpha_2}}}

\preprint{$\hbox to 5 truecm{\hfil SU-GP-93/4-5}
\atop
\hbox to 5 truecm{\hfil gr-qc/9305003}$}
\title{Holomorphic quantum mechanics with a \\
quadratic Hamiltonian constraint}
\author{Jorma Louko\cite{email}}
\address{
Department of Physics, Syracuse University,
Syracuse, New York 13244--1130, USA
}
\date{May 1993}
\maketitle
\begin{abstract}
A finite dimensional system with a quadratic Hamiltonian constraint is
Dirac quantized in holomorphic, antiholomorphic and mixed representations.
A unique inner product is found by imposing Hermitian conjugacy relations on
an operator algebra. The different representations yield drastically different
Hilbert spaces. In particular, all the spaces obtained in the antiholomorphic
representation violate classical expectations for the spectra of certain
operators, whereas no such violation occurs in the holomorphic
representation. A subset of these Hilbert spaces is also recovered in a
configuration space representation. A propagation amplitude obtained from an
(anti)holomorphic path integral is shown to give the matrix elements of the
identity operators in the relevant Hilbert spaces with respect to an
overcomplete basis of representation-dependent generalized coherent states.
Relation to quantization of spatially homogeneous cosmologies is discussed in
view of the no-boundary proposal of Hartle and Hawking and the new variables
of Ashtekar.
\end{abstract}

\narrowtext

\section{Introduction}
\label{sec:intro}

The three standard formulations of quantum theory are Schr\"odinger's wave
function formulation, Heisenberg's operator formulation, and Feynman's
path integral formulation. Although these methods can be considered
equivalent in ordinary nonrelativistic quantum mechanics in their common
domain of validity, there exist systems of interest in which the
applicability and interrelation of these methods is less clear. An example
that has recently received considerable attention is quantum field theory on
spacetimes containing closed timelike loops\cite{time-loops}. Another
example, on which we shall concentrate in this paper, are Hamiltonian systems
whose dynamics is entirely generated by constraints. An important class of
such systems are reparametrization invariant theories, including the general
theory of relativity.

An analogue of the Schr\"odinger method of quantization for constrained
Hamiltonian systems was developed by Dirac\cite{dirac}, the idea being to
promote the (first class) classical constraints into quantum operators that
are postulated to annihilate the wave function. For systems whose dynamics is
entirely generated by constraints this method can be regarded as coinciding
with the Heisenberg method, as such systems possess no preferred, external
time variable. The Dirac method in the form given in Ref.\cite{dirac} is,
however, generally considered incomplete in that it does not specify an inner
product, or any other means of obtaining the probabilistic predictions one
expects of a quantum theory\cite{AAbook,kucharGR13}. Methods for providing
such an inner product have been proposed in
Refs.\cite{AAbook,woodhouse,isham_LH}.

On the other hand, path integral quantization methods have been extensively
applied to constrained Hamiltonian systems, and there is ample evidence
\cite{bata,teitel2,hallhartle/constraint} that appropriately defined path
integrals are related to the constraint equations of Dirac quantization.
Apart from simple exactly solvable examples, such as the relativistic point
particle \cite{henn-teitel,hartle-kuchar}, it appears however unclear what the
relation of such path integrals should be to any Hilbert space structure that
one may have introduced in the Dirac quantization. A recent discussion of
these issues is given in Ref.\cite{hartle_LH}.

In much of the work on path integrals in constrained systems, the attention
has been on (extended) configuration space or phase space path integrals. In
this paper we shall consider a constrained system which admits a natural
class of coherent state, or ``holomorphic," representations in the
Dirac quantization, as well as a natural class of coherent state path
integral constructions. Our first aim is to analyze the quantum theories that
are obtained through the Dirac method in the different coherent state
representations, adopting Ashtekar's suggestion that the inner product be
determined from the Hermitian conjugacy relations on a suitable operator
algebra\cite{AAbook}. Our second aim is to establish that certain coherent
state path integrals do produce quantum amplitudes that can be interpreted
both as operators and as generalized coherent state vectors in the Hilbert
spaces emerging from the Dirac quantization.

Our model consists of two harmonic oscillators with identical frequencies,
with the dynamics given by a single Hamiltonian constraint which sets the
energy difference of the oscillators equal to a prescribed real number.
The constraint is analogous to the Hamiltonian constraint in general
relativity\cite{wald} in that both consist of a sum of a quadratic kinetic
term and a potential term, and further in that the kinetic and potential
terms have indefinite sign. Models of this kind arise from spatially
homogeneous cosmologies that admit a Hamiltonian
formulation\cite{ryan,jantzen12,ash-sam,couss}, and there exists a class of
spatially homogeneous cosmologies that are, modulo certain global caveats,
exactly described by our model\cite{pagescalar,rayrev}.

The classical properties of this and related models were extensively analyzed
in Refs.\cite{schon-haji,haji} from the viewpoint of topological obstructions
to quantization, and difficulties of the Dirac quantization in the position
representation were discussed in\cite{haji}. The Dirac quantization of this
model in Ashtekar's program has been previously considered in
Refs.\cite{AAbook,tate:th} in a representation based on generalized
angular momentum eigenstates.

In the Dirac quantization, we shall consider a class of representations
which generalize the Bargmann representation for a single unconstrained
harmonic oscillator\cite{bargmann,itkuson}. The quantum states are
represented by analytic or antianalytic functions of the ``complex
coordinates" of each oscillator, providing a total of four different
possibilities (analytic in both arguments, antianalytic in both arguments, or
analytic in one argument and antianalytic in the other). The inner product
will be uniquely determined, under certain assumptions, by imposing
Hermitian conjugacy relations on a suitable operator algebra. Unlike in
the case of a single unconstrained harmonic oscillator, all the four
representations will lead into sets of Hilbert spaces. While some of these
sets overlap in part, the intersection of all four sets is empty, and the
individual quantum theories have drastically different properties. In
particular, in all the quantum theories obtained in the representation where
the wave functions are antianalytic in both oscillators, the spectra of
certain operators violate inequalities satisfied by their classical
counterparts. Conversely, no such violation occurs in the representation
where the wave functions are analytic in both oscillators.

We also discuss the Dirac quantization in a position representation in
which the quantum states are represented by functions of the real-valued
configuration space coordinates. In this representation one recovers Hilbert
spaces that are isomorphic to those above mentioned ones in which the
operators do not violate the inequalities satisfied by their classical
counterparts.

Coherent state path integrals are constructed in analogy with those for the
unconstrained harmonic oscillator\cite{itkuson,klauder,klauder-skager}. With
a careful definition, the path integral is shown to yield the matrix elements
of the identity operator in the Hilbert spaces obtained in Dirac's
quantization, with respect to an overcomplete basis of
representation-dependent generalized coherent states. In the representations
where the wave functions are analytic or antianalytic in both arguments, this
basis consists of simultaneous annihilation operator and minimum uncertainty
coherent states associated with the Lie algebra of
$SO(2,1)$\cite{klauder-skager}. In the mixed representations, on the other
hand, the basis consists of displacement operator coherent states associated
with the same Lie algebra\cite{klauder-skager,perelomov}.

Using the relation of our model to spatially homogeneous cosmologies, the
above results are reinterpreted as quantizations of certain cosmological
models\cite{pagescalar,rayrev}. In the position representation, we show that
the ``ground state" wave function in the relevant Hilbert space has the
semiclassical behavior conventionally associated with the no-boundary wave
function of Hartle and Hawking\cite{vatican,HH,hawNB}. The same holds also in
the corresponding coherent state representation. A path integral definition of
the no-boundary wave function beyond the semiclassical estimate remains
however problematic in all the representations.

The paper is organized as follows. In section \ref{sec:model} we introduce
the model and review its classical dynamics in a set of complex phase space
variables. In section \ref{sec:holo} we carry out the Dirac and path integral
quantizations in the representation where the wave functions are analytic in
both arguments, and outline the corresponding results for the representation
where the wave functions are antianalytic in both arguments. Section
\ref{sec:mix} contains the corresponding analysis in the mixed
representations. The Dirac quantization in the position representation is
carried out in section \ref{sec:position}, and the cosmological
interpretation is discussed in section~\ref{sec:cosmology}.

The results are summarized and discussed in section~\ref{sec:conclusions}.
We in particular discuss the relation of our approach to the quantization of
spatially homogeneous cosmologies in the connection representation of
Ashtekar's variables\cite{AAbook}, and to the Born-Oppenheimer expansion in
quantum cosmology. A related model in which the energy difference
constraint is replaced by an energy sum constraint is briefly analyzed in
the appendix.

\section{The model}
\label{sec:model}

We consider a model whose action is
\begin{equation}
S = \int dt
\left(
p_1 {\dot x}_1
+ p_2 {\dot x}_2
- N \cH
\right)
\ \ ,
\label{action}
\end{equation}
where the constraint $\cH$ is given by
\begin{equation}
\cH = \casehalf \!
\left( p_1^2 - p_2^2  + x_1^2 - x_2^2 \right)
-2\delta
\ \ ,
\label{cH}
\end{equation}
and $\delta$ is an arbitrary real number. The unconstrained phase space
$\Gamma$ is ${\bf R}^4$, with global canonical coordinates
$(x_I, p_I)$, $I=1,2$, with Poisson brackets
$\left\{x_I,p_I \right\}=\delta_{IJ}$. $N$ is a Lagrange multiplier
enforcing the constraint.

Physically, the model describes two harmonic oscillators with identical
frequencies and an energy difference equal to~$2\delta$. The classical
equations of motion are easily solved. As a preparation to the quantization,
we shall in this section describe the classical dynamics in terms of
a complex set of functions on the phase space.

To begin, define on $\Gamma$ the complex-valued functions
\begin{equation}
\begin{array}{rcl}
\zi &=& {1\over \sqrt{2}} \left( x_I - i p_I \right) \\
\noalign{\smallskip}
\zbi &=& {1\over \sqrt{2}} \left( x_I + i p_I \right)
\ \ .
\end{array}
\label{zeds}
\end{equation}
The set ${\cal S} = \left\{\zi, \zbi, 1\right\}$ of functions on $\Gamma$ is
closed under the Poisson bracket, $\left\{\zi,\zbj\right\}=i\delta_{IJ}$, and
every sufficiently regular function on $\Gamma$ can be expressed in terms of
(possibly  infinite) sums and products of the elements of~${\cal S}$. The
constraint takes the form
\begin{equation}
\cH = z_1 {\bar z}_1 - z_2 {\bar z}_2
- 2\delta
\ \ .
\label{cHzed}
\end{equation}
The classical Poisson bracket algebra generated by the elements in ${\cal S}$
is therefore sufficiently large for describing the classical dynamics of
the system. This algebra will be used as the starting point of quantization
in the next section.

Next, we need a set of constants of motion. Consider on $\Gamma$ the three
functions
\begin{mathletters}
\label{Jclass}
\begin{eqnarray}
J_+ &=& z_1 z_2 \\
J_- &=& {\bar z}_1 {\bar z}_2 \\
J_0 &=& \casehalf \left(
z_1 {\bar z}_1 +
z_2 {\bar z}_2 \right)
\label{Jclasso}
\end{eqnarray}
\end{mathletters}
whose Poisson brackets form a closed algebra by
\begin{equation}
\begin{array}{rcl}
\left\{J_+,J_-\right\} &=& 2i J_0
\\
\left\{J_0,J_{\pm}\right\} &=& \mp i J_{\pm}
\ \ .
\end{array}
\label{Cso21alg}
\end{equation}
As the $J$'s have (strongly) vanishing Poisson brackets with $\cH$, their
values on the constraint surface $\cH=0$ are constants of motion. It is
straightforward to verify that the $J$'s are a complete set of constants
of motion, in the sense that every classical solution is uniquely specified
by the values of the~$J$'s.

On the constraint surface $\cH=0$, the $J$'s are not algebraically
independent but satisfy the identity
\begin{equation}
-J_0^2 + J_+ J_- =
- \delta^2
\ \ .
\label{CJrel}
\end{equation}
An algebraically independent set of constants of motion that is large enough
to uniquely specify a classical solution is given by the real and imaginary
parts of~$J_\pm$. These two real numbers can take arbitrary values, and they
determine $J_0$ uniquely as the non-negative root of Eq.~(\ref{CJrel}). The
space of constants of motion, denoted by~$\bar\Gamma$, is therefore
topologically~${\bf{R}}^2$.

For $\delta\ne0$ the constraint surface is a manifold, and $\bar\Gamma$
inherits from $\Gamma$ a natural differentiable structure in which a global
coordinate chart is provided by the real and imaginary parts of~$J_\pm$.
The symplectic form $\Omega = \sum_I dp_I \wedge dx_I$ on $\Gamma$ has a
smooth non-degenerate pull-back to~$\bar\Gamma$, given by
\begin{equation}
\Omega_{\bar\Gamma} =
{i \, dJ_+ \wedge dJ_-
\over 2 J_0}
\ \ ,
\end{equation}
where $J_0$ is the positive root of~(\ref{CJrel}). For $\delta=0$ the
constraint surface is not a manifold near the point $x_I=p_I=0$, but there
nevertheless exists a differentiable structure on $\bar\Gamma$ such that the
pull-back of $\Omega$ is smooth and non-degenerate. A global coordinate chart
in this differentiable structure is provided by the real and imaginary parts
of the functions $I_\pm = J_\pm{\left(J_+J_-\right)}^{-1/4}$, and one has
\begin{equation}
\Omega_{\bar\Gamma} =
i \, dI_+ \wedge dI_-
\ \ .
\end{equation}
$\bar\Gamma$ has therefore the structure of a genuine reduced phase space
for all values of~$\delta$.

\section{Quantum theory in the doubly holomorphic representation}
\label{sec:holo}

We now embark on quantization of the model. In subsection
\ref{subsec:can-holo} we apply the Dirac method, following the algebraic
quantization program of Ref.\cite{AAbook}, and choose a representation
analogous to the Bargmann representation of an unconstrained harmonic
oscillator\cite{bargmann,itkuson}. In subsection \ref{subsec:path-holo} we
shall relate the resulting Hilbert spaces to path integral quantization.

\subsection{Doubly holomorphic Dirac quantization}
\label{subsec:can-holo}

We begin by introducing a set of elementary quantum operators,
${\hat{\cal S}}=\left\{\zhi, \zbhi, {\hat {\openone}}\right\}$,
with the commutator algebra
\begin{equation}
\begin{array}{rcl}
\left[\zhi,\zbhj \right] &=& -\delta_{IJ}{\hat{\openone}}
\\
\noalign{\smallskip}
\left[\zhi,{\hat{\openone}} \right] &=&
\left[\zbhi,{\hat{\openone}} \right] = 0
\ \ .
\end{array}
\label{zedcomms}
\end{equation}
Here, and from now on, ${\hat{\openone}}$ stands for the identity operator.
The set ${\hat{\cal S}}$ and its commutator algebra are the quantum
counterparts of the set ${\cal S}$ and its Poisson bracket algebra. The full
quantum operator algebra ${\cal A}$ is the algebra generated
by~${\hat{\cal S}}$.

We wish to represent ${\cal A}$ on a vector space (``space of wave
functions"). We choose the space of complex analytic functions in two
variables,
\begin{equation}
\vholo = \left\{\psi\left(\zi\right) \mid \hbox{$\psi$ analytic in
$\zi$}\right\} \ \ ,
\end{equation}
with the representation
\begin{equation}
\begin{array}{rcl}
\zhi \psi&=& \zi \psi
\\
\noalign{\smallskip}
\zbhi \psi&=& {\displaystyle{\partial \psi \over \partial \zi}}
\ \ .
\end{array}
\label{zedrep1}
\end{equation}
In analogy with the Bargmann representation of the unconstrained harmonic
oscillator\cite{bargmann,itkuson}, we refer to this as the doubly holomorphic
representation. (Note that our conventions for the barred and unbarred
quantities agree with those of Ref.\cite{bargmann} and are opposite of those
of Ref.\cite{itkuson}.) It is not necessary at this stage to be specific
about the allowed singularities in the functions in~$\vholo$. We shall
eventually  consider subspaces of $\vholo$ where the functions will have
precisely defined analyticity properties.

The operator version of the classical Hamiltonian constraint
$\cH$~(\ref{cHzed}) is taken to be
\begin{equation}
\cHh =
{\hat z}_1 {\hat{\bar z}}_1
-{\hat z}_2 {\hat{\bar z}}_2
-2\delta {\hat{\openone}}
\ \ .
\label{cHh}
\end{equation}
This can be thought of as ordering the barred operators to the right of the
unbarred ones. However, as the two oscillators enter $\cH$ with the opposite
signs, any ordering of the form  $\zi\zbi\mapsto (1-t) \zhi\zbhi  + t
\zbhi\zhi$, with the same $t$ for both oscillators, would give the same
result.

The subspace of $\vholo$ where the quantum constraint equation $\cHh\psi=0$
is satisfied is easily found. It is
\begin{equation}
\vholophy=\left\{\psi\in \vholo \mid
\psi\left(\zi\right) = {(z_1/z_2)}^\delta \phi\left(z_1 z_2\right)\right\}
\ \ ,
\label{vholophy}
\end{equation}
where $\phi$ is a complex analytic function of its single argument. Again, it
is not necessary at this stage to be specific about the singularities
in~$\phi$. Note that solving the constraint has introduced no assumptions
about the separability or normalizability of the wave functions.

We now wish to build an algebra of physical operators, that is, a
subalgebra of ${\cal A}$ that would leave $\vholophy$ invariant. To do this,
notice first that the Poisson bracket algebra of the classical $J$'s
(\ref{Jclass}) can be promoted into a commutator algebra by setting
\begin{mathletters}
\label{Jquantum}
\begin{eqnarray}
\jhp &=& {\hat z}_1 {\hat z}_2\\
\jhm &=& {\hat {\bar z}}_1 {\hat {\bar z}}_2\\
\jho &=& \casehalf \!
\left(
{\hat z}_1 {\hat{\bar z}}_1 +
{\hat z}_2 {\hat{\bar z}}_2 + {\hat{\openone}} \right)
\ \ ,
\label{Jquantumo}
\end{eqnarray}
\end{mathletters}
so that the commutators are
\begin{equation}
\begin{array}{rcl}
\left[\jhp,\jhm\right] &=& -2 \jho\\
\noalign{\smallskip}
\left[\jho,\jhpm\right] &=& \pm \jhpm
\ \ .
\end{array}
\label{Qso21alg}
\end{equation}
(Once $\jhp$ and $\jhm$ are chosen as in~(\ref{Jquantum}), the term
proportional to ${\hat{\openone}}$ in $\jho$ is fixed by requiring that the
commutator algebra closes.) Eqs.~(\ref{Qso21alg}) are recognized as the
commutators of the Lie algebra of $SO(2,1)$. As all the $\hat J$'s commute
with $\cHh$, they leave $\vholophy$ invariant. We therefore choose our
physical operator algebra ${\cal A}_{\hbox{\tiny phy}}$ to be the algebra
generated by the set $\left\{\jhpm, \jho, {\hat{\openone}} \right\}$. This
algebra is made into a star-algebra $\astarphy$ by introducing the
star-relation (involution)
\begin{equation}
{\hat J}_{\pm}^{\star} = {\hat J}_{\mp} \ ,
\qquad
{\hat J}_0^{\star} = {\hat J}_0 \ ,
\qquad
{\hat{\openone}}^{\star} = {\hat{\openone}}
\label{Jstar}
\end{equation}
which is inherited from the complex conjugation relations of the
classical~$J$'s.

As the classical $J$'s provide an overcomplete set of constants of motion,
we expect $\astarphy$ to be a sufficiently large algebra for the construction
of a quantum theory. However, in analogy with the algebraic identity
(\ref{CJrel}) satisfied by the $J$'s on the constraint surface, the quantum
${\hat J}$'s on $\vholophy$ are not independent but satisfy the algebraic
relation
\begin{equation}
-{\hat J}_0^2 + \casehalf \left(\jhp\jhm + \jhm\jhp\right) =
\left( \case{1}{4} - \delta^2 \right) {\hat{\openone}}
\ \ .
\label{QJrel}
\end{equation}
The left hand side of Eq.~(\ref{QJrel}) is recognized as the Casimir invariant
of $SO(2,1)$\cite{bargmann-lorentz}. This means that our representation of
the Lie algebra of $SO(2,1)$ on $\vholophy$ contains only those irreducible
representations where the value of the Casimir invariant is
${1\over4}-\delta^2$. The shift from the classical value $-\delta^2$
on the right hand side of Eq.~(\ref{CJrel}) can be traced to our factor
ordering of the constraint and, at a deeper level, to the relation between
reduced phase space quantization and the Dirac
quantization\cite{tate:th,dirac_vs_reduced}.

Writing the vectors in $\vholophy$ as in Eq.~(\ref{vholophy}), in terms of a
function of a single complex variable, the action of the ${\hat J}$'s
takes the form
\begin{mathletters}
\label{jholoact_phi}
\begin{eqnarray}
\left(\jhp \phi\right) (w)  &=&
w \phi (w)
\\
\left(\jhm \phi\right) (w)  &=&
w \phi'' (w) + \phi'(w) - {\delta^2 \over w} \, \phi (w)
\\
\left(\jho\phi\right) (w) &=&
w \phi'(w) + \casehalf \phi(w)
\ \ ,
\end{eqnarray}
\end{mathletters}
where the prime denotes derivative with respect to the argument~$w$. From
this it is clear that the representation of $\astarphy$ on $\vholophy$ is
highly reducible. Consider therefore the subspace where $\phi$ is a linear
combination of arbitrary powers,
\begin{equation}
\vtholophy
= \hbox{Span} \left\{\psi_m\in \vholophy \mid
\psi_m\left(\zi\right) = {(z_1/z_2)}^\delta {(z_1 z_2)}^m, \,
m \in {\bf C} \right\}
\ \ .
\label{vtholophy}
\end{equation}
The action of the ${\hat J}$'s on the basis vectors $\psi_m$ is
\begin{mathletters}
\label{holoactj}
\begin{eqnarray}
\jhp \psi_m &=& \psi_{m+1}
\label{holoactjp}\\
\jhm \psi_m &=& \left( m^2 - \delta^2 \right) \psi_{m-1}
\label{holoactjm}
\\
\jho \psi_m &=& \left(m+\casehalf \right) \psi_m
\ \ ,
\label{holoactjo}
\end{eqnarray}
\end{mathletters}
and $\vtholophy$ therefore carries a representation of~$\astarphy$.
This representation is still reducible, but all the subspaces of $\vtholophy$
carrying irreducible representations are straightforward to find. All
these irreducible representations are cyclic, with some $\psi_m\in\vtholophy$
as the cyclic vector. They have countably infinite dimension, by virtue
of Eq.~(\ref{holoactjp}), and they are all faithful.

(Note that we have not specified a topology on $\vtholophy$ or on its
subspaces. By irreducibility we therefore understand algebraic irreducibility,
which means nonexistence of nontrivial invariant subspaces. It has been
argued that when representing an operator algebra on a Hilbert space, the
physically appropriate criterion is topological irreducibility, which
means nonexistence of nontrivial closed invariant subspaces\cite{rendall}.)

To complete the quantization, we wish to introduce an inner product.
Given a subspace of $\vtholophy$ on which the representation of $\astarphy$
is irreducible, we seek on this subspace an inner product such that the
${\hat J}$'s satisfy the Hermitian conjugacy relations
\begin{equation}
{\hat J}_{\pm}^{\dagger} = {\hat J}_{\mp} \ ,
\qquad
{\hat J}_0^{\dagger} = {\hat J}_0
\label{Jdag}
\end{equation}
which correspond to the star-relation (\ref{Jstar}) of~$\astarphy$. If such
an inner product exists, a Hilbert space is constructed by
Cauchy completion.

It turns out that an inner product satisfying these requirements exists only
for a subset of the irreducible representations; however, when such
an inner product exists, it is unique (up to an overall constant). The proof
of these assertions is straightforward. The Hermiticity of $\jho$ and
Eq.~(\ref{holoactjo}) first tell that the index $m$ is real and the inner
product is diagonal in the basis~$\left\{\psi_m\right\}$. The Hermitian
conjugacy of $\jhpm$ and Eqs.~(\ref{holoactjp})-(\ref{holoactjm}) then relate
the norms of $\psi_m$ for different~$m$, and the requirement that the inner
product be positive definite sets the final restriction on the values
of~$m$. There are just three different cases, which we now list in turn.

For any $\delta\in{\bf R}$, one admissible vector space is
\begin{equation}
\vpholo = \hbox{Span}
\left\{\psi_m\in \vtholophy \mid
m = |\delta| + k,\, k=0,1,2,\ldots\right\}
\ \ .
\label{vpholo}
\end{equation}
The inner product is
\begin{equation}
\left(\psi_m , \psi_{m^\prime} \right) =
\Gamma(m+1+\delta) \Gamma(m+1-\delta)
\delta_{m,m^\prime}
\label{holo_ip}
\end{equation}
where $\delta_{m,m^\prime}$ stands for the Kronecker delta. To examine the
physical content of this space, consider the normal ordered energy
operators for each oscillator,
\begin{equation}
{\hat E}_I = \zhi\zbhi + \casehalf {\hat {\openone}}
\ \ .
\end{equation}
These operators clearly are in~$\astarphy$, and the basis vectors
$\psi_{|\delta| + k}$ are joint eigenvectors of~${\hat E}_I$. The
eigenenergies of the lower energy oscillator (which is the first oscillator
for $\delta<0$ and the second oscillator for $\delta>0$) are $k+\half$, that
is, exactly the positive half-integers that one would anticipate on the basis
of a single unconstrained oscillator. The respective eigenenergies of the
higher energy oscillator are $k+\half+2|\delta|$, which are  half-integers
only if $2\delta$ is an integer. In the degenerate case $\delta=0$ the
eigenenergies of the two oscillators coincide. The spectrum of $\jho$ is
bounded below by $|\delta|+\half$, and this space therefore respects the
classical inequality $J_0\ge|\delta|$.

If $0<|\delta|<\half$, a second admissible vector space is
\begin{equation}
\vmholo = \hbox{Span}
\left\{\psi_m\in \vtholophy \mid
m = -|\delta| + k, \, k=0,1,2,\ldots\right\}
\ \ .
\label{vmholo}
\end{equation}
The inner product is again given by Eq.~(\ref{holo_ip}). Again, the basis
vectors $\psi_{-|\delta|+k}$ are  simultaneous eigenvectors of~${\hat E}_I$.
Now  the eigenenergies of the higher energy oscillator are the
positive half-integers $k+\half$, and those of the lower energy oscillator
are respectively $k+\half-2|\delta|$, which are never half-integers.
The spectrum of $\jho$ is bounded below by $\half-|\delta|$. The classical
inequality $J_0\ge|\delta|$ is therefore violated if
${1\over4}<|\delta|<\half$, but only moderately. Note that the
inequality ${1\over4}<|\delta|<\half$ is precisely the condition under
which the lowest eigenenergy of the lower energy oscillator is negative.

Finally, if $|\delta|<\half$, there exists a one-parameter family of
admissible vector spaces, labeled by a real number $\epsilon$
satisfying $|\delta|<\epsilon<1-|\delta|$. These spaces are
\begin{equation}
\vepsholo = \hbox{Span}
\left\{\psi_m\in \vtholophy \mid
m = \epsilon + k, \, k\in{\bf Z} \right\}
\ \ ,
\end{equation}
and the inner product is again given by Eq.~(\ref{holo_ip}). The spectra of
$\jho$ and ${\hat E}_I$ are unbounded both above and below.

These three cases give the complete list of the inner product spaces. The
respective abstract Hilbert spaces $\hilpholo$, $\hilmholo$ and $\hilepsholo$
are obtained through Cauchy completion. The factorial growth of the inner
product (\ref{holo_ip}) at $m\to\infty$ guarantees that the vectors in
$\hilpmholo$ can be represented by complex analytic functions of the form
given in Eq.~(\ref{vholophy}): the function $\phi(w)$ is such that
$w^{\mp|\delta|}\phi(w)$ (where the upper and lower signs correspond to
$\hilpmholo$, respectively) is analytic everywhere in the finite complex $w$
plane. The spaces $\hilepsholo$, on the other hand, all contain vectors that
cannot be represented by complex analytic functions of the form
given in Eq.~(\ref{vholophy}), the reason being the inverse factorial decay
of the inner product (\ref{holo_ip}) for $m=\epsilon+k$ at $k\to-\infty$.

Therefore, if we require that all vectors in the Hilbert space be
representable by complex analytic functions of the form given in
Eq.~(\ref{vholophy}), only the spaces $\hilpmholo$ remain. The inner product
of any $\psi_{(1)},\psi_{(2)}\in\hilpmholo$ can be written as the
holomorphic integral
\begin{equation}
\left(\psi_{(1)},\psi_{(2)}\right)
=
\int {d {\bar w} \wedge dw \over \pi i}
\,
K_{2\delta} \left( 2 \sqrt{w{\bar w}} \right)
\,
\overline{\phi_{(1)}(w)} \phi_{(2)}(w)
\label{holo_int_ip}
\ \ ,
\end{equation}
where $d {\bar w} \wedge dw = 2i\, \hbox{Re}(w) \wedge \hbox{Im}(w)$,
the integration is over the whole complex $w$ plane, and $K_{2\delta}$
is a modified Bessel function\cite{Grad-Rhyz:K}.

We have thus recovered a set of Hilbert spaces carrying representations of
the $SO(2,1)$ Lie algebra. This set contains a continuum of
spaces in which the spectrum of $\jho$ is unbounded both above and below,
as well as one or two spaces, depending on the value of $\delta$, in which the
spectrum of $\jho$ is bounded below but not above. From the viewpoint of the
$SO(2,1)$ Lie algebra~(\ref{Qso21alg}) alone, this result might seem
surprising. The Lie algebra admits an automorphism which interchanges
$\jhp$ with $\jhm$ and reverses the sign of $\jho$: for any representation
of the algebra, there must therefore exist another representation with an
inverted spectrum of~$\jho$. The reason why our set of representations does
not exhibit such a symmetry is in our choice of the doubly holomorphic
representation. Whereas the algebra (\ref{Qso21alg}) tells that there has to
be a total factor of $\left(m^2-\delta^2\right)$ in the actions of $\jhpm$ in
Eqs.~(\ref{holoactj}), it is the choice of the doubly holomorphic
representation that distributes this factor unsymmetrically between the
raising and lowering operators.

Suppose that, instead of the doubly holomorphic representation, one adopts
a doubly antiholomorphic representation in which the algebra ${\cal{A}}$ is
represented on the vector space
\begin{equation}
\vaholo = \left\{{\tilde\psi}\left(\zbi\right) \mid
\hbox{${\tilde\psi}$ antianalytic in $\zbi$}\right\}
\ \ ,
\end{equation}
by
\begin{equation}
\begin{array}{rcl}
\zhi {\tilde\psi}&=& -
{\displaystyle{\partial {\tilde\psi} \over \partial \zbi}}
\\
\noalign{\smallskip}
\zbhi {\tilde\psi}&=& \zbi {\tilde\psi}
\ \ .
\end{array}
\label{zedrep2}
\end{equation}
The analysis can be carried through in perfect analogy with the above.
The counterpart of Eqs.~(\ref{holoactj}) is now
\begin{mathletters}
\label{antiactj}
\begin{eqnarray}
\jhp {\tilde\psi}_m &=&
\left( {(m+1)}^2 - \delta^2 \right)
{\tilde\psi}_{m+1}
\\
\jhm {\tilde\psi}_m &=&  {\tilde\psi}_{m-1}
\\
\jho {\tilde\psi}_m &=& \left(m+\casehalf \right) {\tilde\psi}_m
\ \ .
\end{eqnarray}
\end{mathletters}
The abstract Hilbert spaces fall again into three cases. There are
counterparts of $\hilpmholo$, with the sign in the spectrum of $\jho$
inverted, and there is a continuum of Hilbert spaces isomorphic
to~$\hilepsholo$. Thus, the spectrum of $\jho$ is unbounded below in {\em
all\/} the Hilbert spaces obtained in the doubly antiholomorphic
representation, and in addition bounded above in the counterparts
of~$\hilpmholo$. This is a drastic violation of the classical range of~$J_0$.
Again, the vectors in the counterparts of $\hilpmholo$ can be represented by
antianalytic functions, whereas the spaces isomorphic to $\hilepsholo$ all
contain vectors that cannot be represented in this way.

Taken together, the set of Hilbert spaces obtained in the doubly
holomorphic representation and the doubly antiholomorphic representation
contains precisely those Hilbert spaces that were obtained in
Refs.\cite{AAbook,tate:th} in the generalized angular momentum
eigenstate representation
\begin{mathletters}
\label{tateactj}
\begin{eqnarray}
\jho |m\rangle &=& \left(m+\casehalf \right)
|m\rangle
\\
\jhp |m\rangle &=&
\sqrt{{(m+1)}^2 - \delta^2} \,
|m+1\rangle
\\
\jhm |m\rangle &=&
\sqrt{m^2 - \delta^2} \,
|m-1\rangle
\ \ ,
\end{eqnarray}
\end{mathletters}
which treats the raising and lowering operators in a manifestly symmetric
fashion.

It is natural to ask what happens in the mixed representations where one of
the oscillators is holomorphic and the other antiholomorphic. This will be
explored in section~\ref{sec:mix}.

\subsection{Doubly holomorphic path-integral quantization}
\label{subsec:path-holo}

We wish now to write a path integral for a ``propagation amplitude"
with boundary conditions such that the amplitude could be related to the
Dirac quantization of subsection~\ref{subsec:can-holo}. For definiteness, we
only consider the doubly holomorphic representation. The results for the
doubly antiholomorphic representation are analogous.

Our starting point is the formal path integral expression
\begin{equation}
G \! \left( \alpha_I ; {\bar \beta}_I \right)
=
\int\limits_{\zbi(t_1)={\bar \beta}_I}^{\zi(t_2)=\alpha_I}
\cD \left( \zi , \zbi , N \right)
\exp\left (i S \right)
\label{holo_pi}
\end{equation}
where the action is given by
\begin{equation}
S = \int_{t_1}^{t_2} dt
\left(
-i \zbi {\dot z}_I
- N \cH
\right)
\ \
- i \zi (t_1) \zbi (t_1)
\ \ .
\label{holo_action}
\end{equation}
A sum over the repeated index $I$ in Eq.~(\ref{holo_action}) is understood.
The quantities to be integrated over are $\zi(t)$, $\zbi(t)$ and $N(t)$, with
the final values of $\zi$ fixed to $\alpha_I$, and the initial values of
$\zbi$ fixed to~${\bar \beta}_I$. The action (\ref{holo_action}) is the
appropriate one for the classical boundary value problem with these boundary
conditions, provided $\zi$ and $\zbi$ are regarded as independent quantities
and not each others' complex conjugates. Through integrations by parts, this
action can be written in a more symmetric fashion with respect to the barred
and unbarred variables\cite{itkuson}.

We wish now to give a meaning to this formal expression.

First, the action (\ref{holo_action}) has a gauge symmetry corresponding to
reparametrizations in~$t$, and this gauge freedom must be eliminated
in the path integral. Adopting the proper-time gauge
${\dot N}=0$\cite{teitel2,henn-teitel}, we obtain
\begin{equation}
G \! \left( \alpha_I ; {\bar \beta}_I \right)
=
(t_2-t_1) \int dN
\int\limits_{\zbi(t_1)={\bar \beta}_I}^{\zi(t_2)=\alpha_I}
\cD \left( \zi , \zbi\right)
\exp\left (i S \right)
\ \ ,
\end{equation}
where $S$ is still
given by Eq.~(\ref{holo_action}) but $N$ is now just a number.

Next, consider the $\cD \left( \zi , \zbi\right)$ integrals. For $N$ real,
the integrand and the boundary conditions are in essence identical to those
in the path integral which would give the conventional coherent state matrix
elements of the time evolution operator for a system of two unconstrained
harmonic oscillators\cite{itkuson,klauder}. We define the measure $\cD
\left( \zi , \zbi\right)$ to be equal to the path measure for two such
unconstrained oscillators, multiplied for later convenience by the numerical
factor $1/(2\pi)$. The result is
\begin{equation}
G \! \left( \alpha_I ; {\bar \beta}_I \right)
=
{1\over 2\pi}
\int dT
\,
\exp \left(
\alpha_1 {\bar \beta}_1 e^{-iT} +
\alpha_2 {\bar \beta}_2 e^{iT}
+ 2 i \delta T
\right)
\ \ ,
\label{Gh:int:T}
\end{equation}
where we have written $(t_2-t_1)N=T$. We take the integrand in
Eq.~(\ref{Gh:int:T}) to be valid also when analytically continued to complex
values of~$T$, and we seek to define the remaining integral over $T$ as a
contour integral in the complex $T$ plane.

Suppose first that $\alpha_I$ and ${\bar \beta}_I$ are all real and
positive. Define in the complex $T$ plane a contour  ${\cal C}_1$ consisting
of three straight lines: a line parallel to the imaginary axis from
$-\pi+i\infty$ to $-\pi$, a line along the real axis from $-\pi$ to $\pi$,
and a line parallel to the imaginary axis from $\pi$ to $\pi+i\infty$.
Similarly, define a contour ${\cal C}_2$ to be the mirror image of ${\cal
C}_1$ with respect to the real axis, going first from $-\pi-i\infty$ to
$-\pi$, then from $-\pi$ to $\pi$, and finally from $\pi$ to $\pi-i\infty$.
The integral in Eq.~(\ref{Gh:int:T}) along each of these contours converges,
with the result\cite{Grad-Rhyz:I}
\begin{equation}
{\left(
\alpha_1 {\bar \beta}_1
\over
\alpha_2 {\bar \beta}_2
\right)} ^\delta
I_{\pm2\delta}
\left( 2 \sqrt{\alpha_1 \alpha_2
{\bar \beta}_1 {\bar \beta}_2 }
\right)
\label{G:temp}
\end{equation}
where $I_{\pm2\delta}$ is the modified Bessel function of the first kind,
and the upper and lower signs refer respectively to ${\cal C}_1$ and~${\cal
C}_2$. If $2\delta$ is an integer, the contributions from  the lines parallel
to the imaginary axis cancel, and both ${\cal C}_1$ and ${\cal C}_2$ are
equivalent to the contour along the real axis from  $-\pi$ to~$\pi$.

The central observation is now that the expression (\ref{G:temp}) defines,
when analytically continued to complex values of $\alpha_I$ and ${\bar
\beta}_I$, quantities that have an interpretation in terms of the Hilbert
spaces $\hilpmholo$. We shall first exhibit this interpretation, and then
discuss in what sense such objects can be seen as arising from the path
integral.

Define the amplitudes $G_\pm$ as
\begin{equation}
G_\pm \!
\left( \alpha_I ; {\bar \beta}_I \right)
=
{\left(
\alpha_1 {\bar \beta}_1
\over
\alpha_2 {\bar \beta}_2
\right)} ^{\!\delta}
I_{\pm2|\delta|}
\left( 2 \sqrt{\alpha_1 \alpha_2
{\bar \beta}_1 {\bar \beta}_2 }
\right)
\ \ ,
\label{G:holo}
\end{equation}
where $G_+$ is defined for all $\delta$, and $G_-$ is defined only in
the case $0<|\delta|<\half$. The arguments $\alpha_I$ and ${\bar \beta}_I$
are complex-valued, and $G_\pm$ is understood as a function on an appropriate
Riemann sheet in each of the arguments. We shall show that $G_\pm$ are the
matrix elements of the identity operator in $\hilpmholo$ in a generalized
coherent state basis.

First, expanding the Bessel function in Eq.~(\ref{G:holo}) as a power series
and using~(\ref{holo_ip}), one obtains
\begin{equation}
G_\pm \!
\left( \alpha_I ; {\bar \beta}_I \right)
=
\sum_m \,
{\psi_m \left(\alpha_I\right)
\overline{\psi_m \left(\beta_I\right)}
\over
\left(\psi_m, \psi_m\right)}
\label{Gholo:expan}
\end{equation}
where the values of the summation index $m$ and the inner product
$\left(\psi_m, \psi_m\right)$ refer to~$\hilpmholo$. Identifying
$\left\{{\overline\psi}_m\right\}$ as a basis of the dual space
of~$\hilpmholo$, one recognizes Eq.~(\ref{Gholo:expan}) as the resolution  of
the identity operator in $\hilpmholo$ with respect to the
basis~$\left\{\psi_m\right\}$. The amplitude $G_\pm$ therefore defines the
identity operator in~$\hilpmholo$.

To see the relation to coherent states, define for fixed ${\bar \beta}_I$ the
functions $\phibetas$ by
\begin{equation}
\phibetas \left( z_I \right)
=
G_\pm \! \left( z_I ; {\bar\beta}_I \right)
\ \ .
\label{phibetas}
\end{equation}
Here $\phibetaplus$ is defined for all ${\bar\beta}_I$ and all~$\delta$,
whereas $\phibetaminus$ is defined only for ${\bar\beta}_1\ne0$ when
$-\half<\delta<0$, and only for ${\bar\beta}_2\ne0$ when $0<\delta<\half$.
{}From Eq.~(\ref{Gholo:expan}) one readily sees that the functions
$\phibetas$ define vectors in~$\hilpmholo$, and further that $\left(
\phibetas , \psi \right) = \psi \left( \beta_I \right)$ for any
$\psi\in\hilpmholo$. In particular,
\begin{equation}
\left( \phialphas , \phibetas \right)
=
G_\pm \! \left( \alpha_I ; {\bar\beta}_I \right)
\ \ .
\end{equation}
Thus, $G_\pm \! \left( \alpha_I ; {\bar\beta}_I \right)$ are the matrix
elements of the identity operator in~$\hilpmholo$ with respect to the states
$\left\{\phibetas\right\}$.

The states $\phibetas$ satisfy
\begin{equation}
\jhm
\phibetas
=
{\bar \beta}_1 {\bar \beta}_2
\phibetas
\ \ ,
\label{annihilation}
\end{equation}
by which they are known as a system of annihilation operator coherent states
associated with the $SO(2,1)$ algebra\cite{klauder-skager}. They are
analogous to conserved charge coherent states, and $\phibetaplus$ in fact
reduces to a conserved charge coherent state when $2\delta$ is an
integer\cite{klauder-skager,nieto}. In particular, $\left\{\phibetas\right\}$
form an overcomplete set in~$\hilpmholo$. The resolution of the identity is
most conveniently written by noticing that apart from a normalization factor,
$\phibetas$ depend on ${\bar \beta}_1$ and ${\bar \beta}_2$ only
through the product~${\bar \beta}_1 {\bar \beta}_2$. One can therefore define
(adopting for the moment Dirac's bra-ket notation) the states
$|{\bar\lambda},\pm\rangle\in\hilpmholo$, labeled by a single complex number
$\bar\lambda$, by
\begin{equation}
\phibetas =
{\left(
{\bar \beta}_1
\over
{\bar \beta}_2
\right)} ^{\!\delta}
|{\bar \beta}_1 {\bar \beta}_2,\pm \rangle
\ \ ,
\end{equation}
where $|{\bar\lambda},+\rangle$ is defined for all~$\bar\lambda$, and
$|{\bar\lambda},-\rangle$ is defined for ${\bar\lambda}\ne0$. Using
Eq.~(\ref{holo_int_ip}), one then obtains
\begin{equation}
{\hat {\openone}} =
\int {d {\bar \lambda} \wedge d\lambda \over \pi i}
\,
K_{2\delta} \left( 2 \sqrt{\lambda{\bar \lambda}} \right)
\,
|{\bar\lambda},\pm\rangle
\langle\lambda,\pm|
\ \ .
\label{holoreso}
\end{equation}

The states $\left\{\phibetas\right\}$ form also a system of minimum
uncertainty coherent states\cite{klauder-skager}. To see this, define the
Hermitian operators
\begin{equation}
\begin{array}{rcl}
\jhx &=&
{\displaystyle{ \half \left( \jhp + \jhm \right) }}
\\
\noalign{\medskip}
\jhy &=&
{\displaystyle{ {1\over 2i} \left( \jhp - \jhm \right)  }}
\ \ ,
\end{array}
\label{Jhxy}
\end{equation}
in terms of which the commutators take the standard form of the $SO(2,1)$
algebra,
\begin{mathletters}
\begin{eqnarray}
\left[ \jhx , \jhy \right] &=& -i \jho
\label{Jxycomm-} \\
\left[ \jhy , \jho \right] &=& i \jhx \\
\left[ \jho , \jhx \right] &=& i \jhy
\ \ .
\end{eqnarray}
\end{mathletters}
{}From~(\ref{Jxycomm-}), the Heisenberg uncertainty relation for $\jhx$ and
$\jhy$ is\cite{klauder-skager}
\begin{equation}
\Bigl\langle{(\Delta \jhx )}^2\Bigr\rangle
\Bigl\langle{(\Delta \jhy )}^2\Bigr\rangle
\ge \case{1}{4}
{\langle \jho \rangle}^2
\ \ .
\label{uncertainty}
\end{equation}
In the state~$\phibetas$, one has by Eq.~(\ref{annihilation})
\begin{equation}
\begin{array}{rcl}
\langle \jhx \rangle &=&
\hbox{Re} \left( \beta_1 \beta_2 \right)
\\
\langle \jhy \rangle &=&
\hbox{Im} \left( \beta_1 \beta_2 \right)
\ \ ,
\end{array}
\end{equation}
and a short computation shows that the uncertainty relation
(\ref{uncertainty}) holds with the equality sign. The interpretation is that
the state $\phibetas$ is as closely as possible peaked around the classical
solution for which $J_+=\beta_1\beta_2$, $J_-={\bar \beta}_1{\bar \beta}_2$.

Finally, let us return to the question of recovering the amplitudes
$G_\pm$ from the path integral for general complex values of the
arguments. Recall that we first defined $G_\pm$ only for positive values of
the arguments by giving a contour prescription in Eq.~(\ref{Gh:int:T}), and
we then analytically continued to complex arguments. It is of interest to ask
whether there is a contour in Eq.~(\ref{Gh:int:T}) that would yield $G_\pm$
directly for all complex values of the arguments. If $2\delta$ is an integer
(in which case only $G_+$ is defined) the answer is yes, and the contour can
be chosen to be along the real $T$ axis from $-\pi$ to~$\pi$. For $2\delta$
not an integer, however, the answer is negative. For any contour such that
Eq.~(\ref{Gh:int:T}) converges for for all $\alpha_I$ and~${\bar\beta}_I$, the
resulting $G$ is a single-valued analytic function in its arguments, and thus
cannot be equal to~$G_\pm$.

It is possible to obtain $G_\pm$ for general $\delta$ and complex arguments
living on the appropriate Riemann sheets if one allows the contour in
Eq.~(\ref{Gh:int:T}) to depend on the arguments. This can be achieved by
contours consisting of two semi-infinite vertical pieces joined by a piece
along the real axis, as in ${\cal C}_1$ and ${\cal C}_2$, but now choosing
the locations of the vertical pieces to depend on the arguments of~$G_\pm$ in
a suitable way. It is however not clear in what sense such a contour
prescription could be justified in terms of the original path
integral~(\ref{holo_pi}).

\section{Quantum theory in the holomorphic-antiholomorphic
representation}
\label{sec:mix}

In this section we shall repeat the analysis of section~\ref{sec:holo}, but
now choosing a mixed representation where the wave functions are analytic
in one oscillator and antianalytic in the other. Subsection
\ref{subsec:can-mix} carries out the Dirac quantization, and subsection
\ref{subsec:path-mix} will construct a propagation amplitude from a path
integral.

\subsection{Holomorphic-antiholomorphic Dirac quantization}
\label{subsec:can-mix}

Our starting point is again the operator algebra ${\cal A}$ defined in
subsection~\ref{subsec:can-holo}. We now represent ${\cal A}$ on
the vector space
\begin{equation}
\vmix =\left\{\chi\left(z_1 , {\bar z}_2 \right) \mid
\hbox{$\chi$ analytic in $z_1$,
antianalytic in ${\bar z}_2$}
\right\}
\label{vmix}
\end{equation}
by
\begin{equation}
\begin{array}{rcl}
{\hat z}_1 \chi&=& z_1 \chi \\
\noalign{\smallskip}
{\hat {\bar z}}_1 \chi &=&
{\displaystyle{\partial \chi \over \partial z_1}}
\\
\noalign{\medskip}
{\hat z}_2 \chi&=& -
{\displaystyle{\partial \chi \over \partial {\bar z}_2}}
\\
{\hat {\bar z}}_2 \chi&=& {\bar z}_2 \chi
\ \ .
\end{array}
\label{zedrep:mix}
\end{equation}
We refer to this as the holomorphic-antiholomorphic representation.
With the quantum constraint $\cHh$ still given by Eq.~(\ref{cHh}), the space
of solutions to the constraint equation $\cHh\chi=0$ is
\begin{equation}
\vmixphy=\left\{\chi\in \vmix \mid
\chi\left(z_1 , {\bar z}_2 \right) =
{(z_1{\bar z}_2)}^{\delta-(1/2)}
\rho\left(z_1/{\bar z}_2\right)\right\}
\ \ ,
\label{vmixphy}
\end{equation}
where $\rho$ is a complex analytic function of its argument. Again, we
defer the question of the singularities in $\rho$ until later.

The star-algebra $\astarphy$ of physical operators is as in
section~\ref{sec:holo}. Writing a vector in $\vmixphy$ as
in Eq.~(\ref{vmixphy}), the action of the ${\hat J}$'s is
\begin{mathletters}
\label{jmixact_phi}
\begin{eqnarray}
\left(\jhp \rho\right) (w)  &=&
- \left( \delta - \casehalf \right) w \rho(w) +
w^2 \rho'(w)
\\
\left(\jhm \rho\right) (w)  &=&
{\left( \delta - \casehalf \right) \over w} \, \rho(w)
+ \rho'(w)
\\
\left(\jho\rho\right) (w) &=&
w \rho'(w)
\ \ ,
\end{eqnarray}
\end{mathletters}
where the prime denotes derivative with respect to the argument. This
representation is again highly reducible. Consider therefore the subspace
where $\rho$ is a linear combination of arbitrary powers,
\begin{equation}
\vtmixphy
= \hbox{Span} \left\{\chi_m\in \vmixphy \mid
\chi_m \left( z_1, {\bar z}_2 \right) =
{(z_1{\bar z}_2)}^{\delta-(1/2)}
{\left(z_1/{\bar z}_2\right)}^{m+(1/2)}, \,
m \in {\bf C} \right\}
\ \ .
\label{vtmixphy}
\end{equation}
The action of the ${\hat J}$'s on the basis vectors $\chi_m$ is
\begin{mathletters}
\label{mixactj}
\begin{eqnarray}
\jhp \chi_m &=&
\left( m+1 - \delta \right) \chi_{m+1}
\label{mixactjp}\\
\jhm \chi_m &=&
\left( m + \delta \right) \chi_{m-1}
\label{mixactjm}\\
\jho \chi_m &=& \left(m+\casehalf \right) \chi_m
\ \ ,
\label{mixactjo}
\end{eqnarray}
\end{mathletters}
and $\vtmixphy$ therefore carries a representation of~$\astarphy$. This
representation is still reducible, but the subspaces carrying irreducible
representations are straightforward to find. All these irreducible
representations are cyclic, with some $\chi_m\in\vtmixphy$ as the cyclic
vector. Representations with countably infinite dimension exist for any
$\delta$, whereas finite dimensional representations exist only when
$2\delta$ is a positive integer.

We seek those irreducible representations that admit an inner
product satisfying the Hermitian conjugacy relations~(\ref{Jdag}). Again,
such an inner product exists only for a subset of the irreducible
representations, but when it exists, it is unique (up to an overall
constant). There are just three different cases.

For $\delta<\half$, one admissible vector space is
\begin{equation}
\vonemix = \hbox{Span}
\left\{\chi_m\in \vtmixphy \mid
m = -\delta + k,\, k=0,1,2,\ldots\right\}
\ \ .
\label{vonemix}
\end{equation}
The inner product is
\begin{equation}
\left(\chi_m , \chi_{m^\prime} \right) =
{\Gamma(m+1+\delta) \over \Gamma(m+1-\delta)}
\,
\delta_{m,m^\prime}
\ \ ,
\label{mix_ip1}
\end{equation}
where $\delta_{m,m^\prime}$ stands for the Kronecker delta. For $\delta\le0$,
the representation of $\astarphy$ on $\vonemix$ is isomorphic to that
on $\vpholo$ in subsection~\ref{subsec:can-holo}, and for
$0<\delta<\half$ it is isomorphic to that on~$\vmholo$. In terms of the basis
vectors, the isomorphism is
\begin{equation}
\chi_{-\delta+k} \longmapsto
{1 \over \Gamma(k+1-2\delta)} \,
\psi_{-\delta+k}
\ \ .
\end{equation}
The eigenenergies of the first oscillator are now the positive
half-integers $k+\half$, and those of the second oscillator are respectively
$k+\half-2\delta$.

For $\delta<\half$, a second admissible vector space is
\begin{equation}
\vtwomix = \hbox{Span}
\left\{\chi_m\in \vtmixphy \mid
m = \delta-1 - k, \, k=0,1,2,\ldots\right\}
\ \ .
\label{vtwomix}
\end{equation}
The inner product is
\begin{equation}
\left(\chi_m , \chi_{m^\prime} \right) =
{\Gamma(\delta-m) \over \Gamma(-\delta-m)}
\,
\delta_{m,m^\prime}
\ \ .
\label{mix_ip2}
\end{equation}
The eigenenergies of the second oscillator are $-k-\half$, and those of the
first oscillator respectively $2\delta-k-\half$.  In particular, the spectrum
of $\jho$ is bounded above by $\delta-\half$, and thus negative definite.
The representation of $\astarphy$ on $\vtwomix$ is therefore not isomorphic
to any of the representations obtained in the doubly holomorphic
representation. Instead, it is isomorphic to a representation on the doubly
antiholomorphic counterpart of $\vpholo$ or~$\vmholo$, depending on the
value of~$\delta$, as was discussed at the end of
subsection~\ref{subsec:can-holo}.

Finally, if $|\delta|<\half$, there exists a one-parameter family of
admissible vector spaces, labeled by a real number $\epsilon$ satisfying
$|\delta|<\epsilon<1-|\delta|$. These spaces are
\begin{equation}
\vepsmix = \hbox{Span}
\left\{\chi_m\in \vtmixphy \mid
m = \epsilon + k, \, k\in{\bf Z} \right\}
\ \ ,
\end{equation}
and the inner product is given by Eq.~(\ref{mix_ip1}). These representations
are isomorphic to those obtained in subsection \ref{subsec:can-holo}
on~$\vepsholo$, with the isomorphism
\begin{equation}
\chi_m \longmapsto
{1 \over \Gamma(m+1-\delta)} \,
\psi_m
\ \ .
\end{equation}

These three cases give the complete list of the inner product spaces.
The respective abstract Hilbert spaces $\hilonemix$, $\hiltwomix$ and
$\hilepsmix$ are obtained by Cauchy completion. Vectors in $\hilonemix$ and
$\hiltwomix$ can be represented by complex analytic/antianalytic functions
of the form given in Eq.~(\ref{vmixphy}). In $\hilonemix$,
$w^{\delta-(1/2)}\rho(w)$ is analytic in the open unit disk $|w|<1$. In
$\hiltwomix$, $w^{(1/2)-\delta}\rho(w)$ is analytic in the domain $|w|>1$ in
the compactified complex plane, including the point $w=\infty$. The spaces
$\hilepsmix$, on the other hand, all contain vectors that cannot be
represented by functions of the form given in Eq.~(\ref{vmixphy}), the reason
being the  divergence of the infinite series expression for $\rho(w)$ for
$|w|\ne1$.

Therefore, if we require that all vectors in the Hilbert space be
representable by functions of the form given in Eq.~(\ref{vmixphy}), only the
spaces $\hilonemix$ and $\hiltwomix$ remain. For $\delta<0$, the inner
products in these two spaces can be expressed as holomorphic
integrals. A formula valid for any $\chi_{(1)},\chi_{(2)}\in\hilonemix$ is
\begin{equation}
\left(\chi_{(1)},\chi_{(2)}\right)
=
{1\over \Gamma(-2\delta)}
\int_{|w|<1}
{d {\bar w} \wedge dw \over 2\pi i}
\,
{\left( w{\bar w} \right)}^{\delta-(1/2)}
{\left(1-w{\bar w}\right)}^{-2\delta-1}
\,
\overline{\rho_{(1)}(w)} \rho_{(2)}(w)
\label{mix_int_ip1}
\ \ .
\end{equation}
For $\hiltwomix$ an analogous formula holds with the integration domain
${|w|>1}$.

We have thus recovered a set of Hilbert spaces carrying representations
of~$\astarphy$. From the discussion at the end of subsection
\ref{subsec:can-holo} we see that this set does not contain all the
Hilbert spaces that were obtained in Refs.\cite{AAbook,tate:th} in
the symmetric representation~(\ref{tateactj}). The reason is that the choice
of the holomorphic-antiholomorphic representation has introduced an
asymmetry, this time between the two oscillators: the representation
(\ref{mixactj}) is not invariant under reversing the sign of~$\delta$. In
particular, the holomorphic-antiholomorphic representation leads to
Hilbert spaces only for $\delta<\half$.

Suppose that, instead of the holomorphic-antiholomorphic representation,
one starts with the antiholomorphic-holomorphic representation where
the two oscillators in Eqs.\ (\ref{vmix}) and (\ref{zedrep:mix}) are
interchanged. The analysis proceeds in exact parallel to the above. The
counterpart of Eqs.~(\ref{mixactj}) will have the inverse sign in $\delta$,
and one arrives at counterparts of $\hilonemix$ and $\hiltwomix$ where the
oscillators are interchanged, as well as at a set of Hilbert spaces
isomorphic to~$\hilepsmix$. Taken together, the two mixed representations
therefore recover all the Hilbert spaces that were obtained in the
representation~(\ref{tateactj}).

\subsection{Holomorphic-antiholomorphic path-integral quantization}
\label{subsec:path-mix}

We wish to write a path integral which could be related to the Hilbert
spaces obtained in subsection~\ref{subsec:can-mix}. For definiteness, we
only consider the holomorphic-antiholomorphic representation, and we assume
$\delta<\half$ throughout the subsection. Analogous results hold in the case
$\delta>-\half$ for the antiholomorphic-holomorphic representation.

Our starting point is now the expression
\begin{equation}
G \! \left( \alpha_1, {\bar\alpha}_2 ; {\bar\beta}_1, \beta_2 \right)
=
\int\limits_{\textstyle{
{\bar z}_1(t_1)={\bar \beta}_1
\atop
z_2(t_1)=\beta_2}}
^{\textstyle{
z_1(t_2)=\alpha_1
\atop
{\bar z}_2(t_2)={\bar\alpha}_2}}
\cD \left( \zi , \zbi , N \right)
\exp\left (i S \right)
\label{mix_pi}
\end{equation}
with the action
\begin{equation}
S = \int_{t_1}^{t_2} dt
\left(
-i \zbi {\dot z}_I
- N \cH
\right)
\ \
- i z_1 (t_1) {\bar z}_1 (t_1)
+ i z_2 (t_2) {\bar z}_2 (t_2)
\ \ .
\label{mix_action}
\end{equation}
A sum over the repeated index $I$ in Eq.~(\ref{mix_action}) is understood.
The boundary conditions in the integral consist of fixing the final values of
$z_1$ and ${\bar z}_2$ respectively to $\alpha_1$ and ${\bar\alpha}_2$, and
the initial values of ${\bar z}_1$ and $z_2$ respectively to
${\bar\beta}_1$ and~$\beta_2$. The action (\ref{mix_action}) is the
appropriate one for the classical boundary value problem with these
boundary conditions, provided $\zi$ and $\zbi$ are considered independent
quantities.

To give a meaning to this formal expression, we again first adopt the proper
time gauge ${\dot N}=0$. The result is
\begin{equation}
G \! \left( \alpha_1, {\bar\alpha}_2 ; {\bar\beta}_1, \beta_2 \right)
=
(t_2-t_1) \int dN
\int\limits_{\textstyle{
{\bar z}_1(t_1)={\bar \beta}_1
\atop
z_2(t_1)=\beta_2}}
^{\textstyle{
z_1(t_2)=\alpha_1
\atop
{\bar z}_2(t_2)={\bar\alpha}_2}}
\cD \left( \zi , \zbi\right)
\exp\left (i S \right)
\ \ .
\label{Gmix:gauge}
\end{equation}
Consider then the $\cD \left( \zi , \zbi\right)$ integrals. The path integral
for the first oscillator is formally identical to that in
section~\ref{sec:holo}. The path integral for the second oscillator is
formally closely similar. However, to define the integral in a consistent
manner, we now need a more careful discussion of the factor ordering.

Recall that for a single unconstrained oscillator, the usual holomorphic path
integral gives just the exponential of the classical action with the
appropriate boundary data\cite{itkuson}. More precisely, the classical action
is evaluated with the Hamiltonian~$z{\bar z}$, and the result is the coherent
state matrix element of the time evolution operator for the normal
ordered quantum Hamiltonian~${\hat{z}}{\hat{\bar{z}}}$. In other words, the
classical Hamiltonian $z{\bar z}$ corresponds to the quantum Hamiltonian in
which the operator represented by differentiation gets ordered to the right.
This was the definition adopted in subsection \ref{subsec:path-holo} for both
oscillators, and we shall again adopt this definition here for the first
oscillator. For the second oscillator, on the other hand, we would like to
define the path integral in Eq.~(\ref{Gmix:gauge}) so as to correspond to the
operator ordering~$-{\hat z}_2 {\hat{\bar z}}_2$, in which the operator
represented by differentiation is now on the left. This suggests that the path
integral for the second oscillator should again be given by the exponential
of the classical action, but now evaluated with the classical Hamiltonian
$1-z_2{\bar z}_2$ rather than just~$-z_2{\bar  z}_2$. We shall adopt this
definition. It will be shown that the resulting amplitude can be interpreted
in terms of the spaces $\hilonemix$ and $\hiltwomix$ of
subsection~\ref{subsec:can-mix}.

With the above definitions, the amplitude takes the form
\begin{equation}
G \! \left( \alpha_1, {\bar\alpha}_2 ; {\bar\beta}_1, \beta_2 \right)
=
-i \int dT
\,
\exp \left[
\left( \alpha_1 {\bar \beta}_1
- {\bar\alpha}_2 \beta_2 \right)
e^{-iT}
+ i (2\delta-1) T
\right]
\ \ .
\label{Gm:int:T}
\end{equation}
We have written $(t_2-t_1)N=T$, and the integrand is taken to be valid in all
of the complex $T$ plane. For convenience, the measure has been chosen to
include a numerical factor of~$-i$.

The remaining integral over $T$ is defined via complex integration and
analytic continuation in the same fashion as in
subsection~\ref{subsec:path-holo}. For example, in the case ${\bar\alpha}_2
\beta_2 -\alpha_1 {\bar \beta}_1>0$ the $T$ contour can be taken along the
positive imaginary axis. The result is
\begin{equation}
G \! \left( \alpha_1, {\bar\alpha}_2 ; {\bar\beta}_1, \beta_2 \right)
=
{\Gamma(1-2\delta) \over
{\left(
{\bar\alpha}_2 \beta_2 -\alpha_1 {\bar \beta}_1
\right)}^{1-2\delta}}
\ \ .
\label{G:mix}
\end{equation}
We understand $G \! \left( \alpha_1, {\bar\alpha}_2 ; {\bar\beta}_1,
\beta_2 \right)$ as a complex analytic function on an appropriate Riemann
sheet in each of its arguments. We shall show that $G \! \left( \alpha_1,
{\bar\alpha}_2 ; {\bar\beta}_1, \beta_2 \right)$ are the matrix elements of
the identity operator in $\hilonemix$ and $\hiltwomix$ in a generalized
coherent state basis which differs from the one discussed in
subsection~\ref{subsec:path-holo}.

Suppose first that the inequalities
\begin{mathletters}
\label{ineq1}
\begin{eqnarray}
|\alpha_1/{\bar\alpha}_2| &<& 1\\
|{\bar\beta}_1/\beta_2| &<& 1
\label{ineq1b}
\end{eqnarray}
\end{mathletters}
hold. Expanding $G$ in the quantity
$\alpha_1{\bar\beta}_1/{\bar\alpha}_2\beta_2$ by the binomial
theorem and using Eq.~(\ref{mix_ip1}), one has
\begin{equation}
G \! \left(
\alpha_1, {\bar\alpha}_2 ; {\bar\beta}_1, \beta_2 \right) =
\sum_m \,
{\chi_m \left(\alpha_1, {\bar\alpha}_2\right)
\overline{\chi_m \left(\beta_1, {\bar\beta}_2\right)}
\over
\left(\chi_m, \chi_m\right)}
\ \ ,
\label{Gmix:expan1}
\end{equation}
where the values of the summation index $m$ and the inner product
$\left(\chi_m, \chi_m\right)$ refer to~$\hilonemix$. Identifying
$\left\{{\bar\chi}_m\right\}$ as as basis of the dual space of~$\hilonemix$,
one recognizes Eq.~(\ref{Gmix:expan1}) as the resolution of the identity
operator in $\hilonemix$ with respect to the basis~$\left\{\chi_m\right\}$.
The amplitude $G$ therefore defines the identity operator in~$\hilonemix$.

To see the relation to coherent states, still assuming Eqs.~(\ref{ineq1}) to
hold (so that in particular $\beta_2\ne0$), define for fixed ${\bar\beta}_1$
and $\beta_2$ the function $\psibetas$ by
\begin{equation}
\psibetas \left(z_1,{\bar z}_2\right)
=
G \! \left( z_1, {\bar z}_2 ; {\bar\beta}_1, \beta_2 \right)
\ \ .
\label{psibetas}
\end{equation}
{}From Eq.~(\ref{Gmix:expan1}) one sees that that $\psibetas$ defines a vector
in~$\hilonemix$. (Note that this would not hold if the condition
(\ref{ineq1b}) were relaxed.) Further, $\left( \psibetas , \chi \right) =
\chi \left( \beta_1, {\bar \beta}_2 \right)$ for any $\chi\in\hilonemix$.
In particular,
\begin{equation}
\left(
\psialphas , \psibetas \right)
=
G \! \left(
\alpha_1, {\bar\alpha}_2 ; {\bar\beta}_1, \beta_2 \right)
\ \ .
\label{Psi_product}
\end{equation}
Thus, $G \! \left(\alpha_1, {\bar\alpha}_2 ; {\bar\beta}_1, \beta_2 \right)$
are the matrix elements of the identity operator in~$\hilonemix$ with respect
to the states $\left\{\psibetas\right\}$.

A direct computation shows that $\psibetas$ can be obtained by operating on
the state $\chi_{-\delta}$ annihilated by $\jhm$ with the exponential of the
raising operator,
\begin{equation}
\psibetas =
\Gamma(1-2\delta) {(\beta_2)}^{2\delta-1}
\exp\left[ \left({\bar\beta}_1/\beta_2\right) \jhp \right]
\chi_{-\delta}
\ \ .
\label{displacement1}
\end{equation}
This means that $\left\{\psibetas\right\}$ are a system of displacement
operator coherent states associated with the $SO(2,1)$
algebra\cite{klauder-skager,perelomov}, and  from the construction it is
clear that they are an overcomplete set in~$\hilonemix$. For $\delta<0$, one
can use Eq.~(\ref{mix_int_ip1}) to write the resolution of the identity in
terms of the states $\left\{\psibetas\right\}$ in a form analogous to
Eq.~(\ref{holoreso}).

Note that $\psibetas$ is a lowering operator coherent state only in the
trivial case ${\bar\beta}_1=0$.  Also, it can be verified that $\psibetas$ is
a minimum uncertainty coherent state for $\jhx$ and $\jhy$ only when
${\bar\beta}_1/\beta_2$ is either purely real or purely imaginary.

Suppose then that the inequalities (\ref{ineq1}) are reversed. In this case
it is seen in an analogous fashion that $G$ defines, up to a
phase~${(-1)}^{2\delta-1}$, the identity operator in~$\hiltwomix$. The
functions $\psibetas$ (\ref{psibetas}) again define displacement operator
coherent states in~$\hiltwomix$, but now with respect to the lowering
operator,
\begin{equation}
\psibetas =
\Gamma(1-2\delta) {(-{\bar\beta}_1)}^{2\delta-1}
\exp\left[ -\left(\beta_2/{\bar\beta}_1\right) \jhm \right]
\chi_{\delta-1}
\ \ .
\label{displacement2}
\end{equation}
Again, $G \! \left( \alpha_1, {\bar\alpha}_2 ; {\bar\beta}_1, \beta_2
\right)$ are (up to the phase~${(-1)}^{2\delta-1}$) the matrix elements of
the identity operator in~$\hiltwomix$ with respect to the states
$\left\{\psibetas\right\}$.

\section{Dirac quantization in the position representation}
\label{sec:position}

In this section we shall carry out the Dirac quantization in a position
representation in which the state vectors are functions of~$x_I$. The
steps are closely similar to those in the previous sections.

We now represent the operator algebra ${\cal A}$ on the space of smooth
functions of two real variables,
\begin{equation}
\vpos = \left\{\xi\left(x_1,x_2\right) \mid
\hbox{$\xi\in C^\infty\left({\bf R}^2\right)$}\right\}
\ \ .
\end{equation}
$\zhi$ and $\zbhi$ are taken to act as the usual creation and annihilation
operators in the position representation of unconstrained oscillators,
\begin{equation}
\begin{array}{rcl}
\zhi \xi&=&
{\displaystyle{
{1\over \sqrt{2}} \left( x_I -
{\partial\over\partial x_I} \right)
\xi }}
\\
\noalign{\smallskip}
\zbhi \xi&=&
{\displaystyle{
{1\over \sqrt{2}} \left( x_I +
{\partial\over\partial x_I} \right)
\xi }}
\ \ .
\end{array}
\label{posrep}
\end{equation}

With the Hamiltonian constraint $\cHh$~(\ref{cHh}), the space of all
solutions to the constraint equation $\cHh\xi=0$ does not have a simple
characterization as in the previous sections.  We therefore immediately
restrict the attention to the space spanned by solutions that are separable
in $x_1$ and~$x_2$. This space, denoted by~$\vtposphy$, can be
expressed as a direct sum of four subspaces,
\begin{equation}
\vtposphy =
V_{UU} \oplus  V_{UV} \oplus
V_{VU} \oplus
V_{VV}
\ \ ,
\end{equation}
such that
\begin{equation}
\begin{array}{rcl}
V_{UU} &=& \hbox{Span}
\left\{\xi_{UU,m} \in \vpos \mid m\in {\bf C}\right\}
\\
V_{UV} &=& \hbox{Span}
\left\{\xi_{UV,m} \in \vpos \mid m\in {\bf C}\right\}
\\
V_{VU} &=& \hbox{Span}
\left\{\xi_{VU,m} \in \vpos \mid m\in {\bf C}\right\}
\\
V_{VV} &=& \hbox{Span}
\left\{\xi_{VV,m} \in \vpos \mid m\in {\bf C}\right\}
\ \ ,
\end{array}
\label{four_spaces}
\end{equation}
where
\begin{equation}
\begin{array}{rcl}
\xi_{UU,m} \left(x_1,x_2\right)
&=&
U\!\left(-\delta-m-\casehalf, \sqrt{2}x_1 \right)
U\!\left(\delta-m-\casehalf, \sqrt{2}x_2 \right)
\\
\noalign{\smallskip}
\xi_{UV,m} \left(x_1,x_2\right)
&=&
U\!\left(-\delta-m-\casehalf, \sqrt{2}x_1 \right)
V\!\left(\delta-m-\casehalf, \sqrt{2}x_2 \right)
\\
\noalign{\smallskip}
\xi_{VU,m} \left(x_1,x_2\right)
&=&
V\!\left(-\delta-m-\casehalf, \sqrt{2}x_1 \right)
U\!\left(\delta-m-\casehalf, \sqrt{2}x_2 \right)
\\
\noalign{\smallskip}
\xi_{VV,m} \left(x_1,x_2\right)
&=&
V\!\left(-\delta-m-\casehalf, \sqrt{2}x_1 \right)
V\!\left(\delta-m-\casehalf, \sqrt{2}x_2 \right)
\ \ .
\end{array}
\end{equation}
Here $U$ and $V$ are the two independent solutions to the parabolic cylinder
equation. Their properties, including the behavior under the action of
$\zhi$ and $\zbhi$~(\ref{posrep}), can be found in Ref.\cite{Ab-Steg}.

All the four spaces in Eqs.~(\ref{four_spaces}) are invariant under the
action of~$\astarphy$. The representation of $\astarphy$ on $V_{UU}$ is
isomorphic to that on $\vtholophy$~(\ref{vtholophy}) encountered in
section~\ref{sec:holo}, with the isomorphism $\xi_{UU,m}\mapsto\psi_m$. The
representation of $\astarphy$ on $V_{UV}$ is isomorphic to that on
$\vtmixphy$~(\ref{vtmixphy}) encountered in section~\ref{sec:mix}, with the
isomorphism  $\xi_{UV,m}\mapsto\chi_m$. Similarly, the representation on
$V_{VV}$ is isomorphic to that given by Eqs.~(\ref{antiactj}) on the doubly
antiholomorphic counterpart of $\vtholophy$, and the representation on
$V_{VU}$ is isomorphic to that on the antiholomorphic-holomorphic
counterpart of $\vtmixphy$.

We therefore see that all the irreducible representations of $\astarphy$ that
were obtained in the doubly holomorphic, doubly antiholomorphic, and mixed
representations are isomorphic to representations of $\astarphy$ on
linear subspaces of~$\vtposphy$. Conversely, all irreducible representations
of $\astarphy$ on linear subspaces of $\vtposphy$ are isomorphic to
representations that were obtained in the doubly holomorphic, doubly
antiholomorphic, or mixed representations. Thus, the imposition of the
Hermitian conjugacy relations (\ref{Jdag}) for an inner product on subspaces
of $\vtposphy$ proceeds exactly as before, and the resulting abstract Hilbert
spaces are isomorphic to precisely those that were obtained in the previous
sections. In this sense, the position representation built on solutions to
the constraint equation that are separable in $x_1$ and~$x_2$ is equivalent
to the generalized angular momentum eigenstate representation (\ref{tateactj})
adopted in Refs.\cite{AAbook,tate:th}.

After the Cauchy completion, many of the Hilbert spaces will contain vectors
that are not representable by functions of~$x_I$. However, on certain
subspaces of $\vtposphy$ on which the representations of $\astarphy$ are
isomorphic to those on~$\vpmholo$, the inner product can be written as an
integral formula which guarantees via standard $L^2$ theory that the
vectors in the completion are representable by functions of~$x_I$.  We shall
now demonstrate this. For concreteness, we only deal explicitly with the case
in which the eigenenergies of the first oscillator are the positive
half-integers. This means that we assume $\delta<\half$, and the Hilbert
spaces are isomorphic to~$\hilonemix$. The case in which the
eigenenergies of the second oscillator are the positive half-integers and the
Hilbert spaces are isomorphic to~$\hiltwomix$ is analogous.

Let $A$ and $B$ be given complex numbers, not both equal to zero. Assuming
$\delta<\half$, consider the space $\vonepos\subset\vtposphy$ given by
\begin{equation}
\vonepos = \hbox{Span}
\left\{\xi_k\in \vtposphy \mid
k=0,1,2,\ldots\right\}
\;\subset\;
V_{UU} \oplus  V_{UV}
\ \ ,
\label{vonepos}
\end{equation}
where
\begin{eqnarray}
\xi_k\left(x_1, x_2\right) = &&
U\!\left(-k-\casehalf, \sqrt{2}x_1 \right)
\nonumber \\
&& \times \left[
{A\over \Gamma(k+1-2\delta)}\, U\!\left(2\delta-k-\casehalf, \sqrt{2}x_2
\right) + B \,
V\!\left(2\delta-k-\casehalf, \sqrt{2}x_2 \right)
\right]
\ \ .
\label{xi}
\end{eqnarray}
Here $U$ and $V$ are the parabolic cylinder functions as above. The
representation of $\astarphy$ on $\vonepos$ is isomorphic to that on
$\vonemix$~(\ref{vonemix}), with the isomorphism  $\xi_k \mapsto
\chi_{-\delta+k}$.  From Eq.~(\ref{mix_ip1}), the corresponding inner product
on $\vonepos$ is
\begin{equation}
\left(\xi_k , \xi_{k^\prime} \right) =
{k! \over \Gamma(k+1-2\delta)} \,
\delta_{k,k^\prime}
\ \ .
\label{pos_ip}
\end{equation}
We wish to demonstrate that, for certain values of $A$ and $B$, this inner
product can be written as an integral involving~$x_I$.

Consider first the special case where $2\delta$ is an integer; by assumption
it is then negative or zero. We take $B=0$ and $A=(1/\sqrt\pi)$. Now, $\xi_k$
reduce to products of (unconventionally normalized) harmonic oscillator
eigenfunctions, and the inner product can be written as
\begin{equation}
\left(\xi_k , \xi_{k^\prime} \right) =
\int dx_1 dx_2 \,
\overline{\xi_k (x_1,x_2)} \, \xi_{k^\prime} (x_1,x_2)
\ \ .
\label{pos_ip1}
\end{equation}
Upon the Cauchy completion one therefore recovers a subspace of the
conventional $L^2\left({\bf R}^2\right)$ Hilbert space of two unconstrained
harmonic oscillators. The role of the constraint has been just to choose the
Hilbert subspace where the energy difference is equal to the
integer~$2\delta$.

Consider then the case of general $\delta<\half$. As the quantum constraint
in the position representation is of the Klein-Gordon form, a natural
candidate for the inner product is
\begin{equation}
\left(\xi_k , \xi_{k^\prime} \right) =
i \int dx_1 \,
\overline{\xi_k (x_1,x_2)}
\> \tensor{\partial}_2 \,
\xi_{k^\prime} (x_1,x_2)
\ \ ,
\label{pos_ip2}
\end{equation}
where $\tensor{\partial}_2 = \roarrow{\partial}_{x_2} -
\loarrow{\partial}_{x_2}$ as usual.  From the Wronskian relation of the
parabolic cylinder functions\cite{Ab-Steg} it follows that (\ref{pos_ip2})
holds, provided we set
\begin{equation}
A {\bar B} - B {\bar A} = {i \over 2}
\ \ .
\label{ABcondition}
\end{equation}
Upon the Cauchy completion one thus arrives at a one-particle Klein-Gordon
theory in two-dimensional Minkowski space with an external potential. The
constants $A$ and $B$ satisfying the condition~(\ref{ABcondition}) specify
the split between the positive and negative frequencies, and the Hilbert
space is the one-particle Hilbert space associated with the positive frequency
solutions. The integral kernel of the identity operator in this Hilbert space
is the positive frequency Wightman function,
\begin{equation}
G^+\left(x_1, x_2 ; x^\prime_1,
x^\prime_2\right) =
\sum_k
{\xi_k\left(x_1, x_2\right)
\overline{\xi_k\left( x^\prime_1, x^\prime_2\right)}
\over
\left(\xi_k , \xi_k \right)}
\ \ ,
\end{equation}
which satisfies
\begin{equation}
\xi \left(x_1, x_2\right) =
i \int dx^\prime_1 \,
G^+\left(x_1, x_2 ; x^\prime_1, x^\prime_2\right)
\tensor{\partial}_{2^\prime}
\xi \left( x^\prime_1, x^\prime_2\right)
\ \ .
\end{equation}

By construction, all choices for the constants $A$ and $B$ satisfying
Eq.~(\ref{ABcondition}) lead to isomorphic quantum theories, in spite of the
two real parameters (in addition to the overall phase) contained in these
constants. However, if one interprets this theory as the positive frequency
single-particle sector of a many-particle Klein-Gordon theory, one would
usually take the viewpoint that not all the information of interest in the
many-particle theory is contained in the operators in~$\astarphy$. One could,
for example, study the responses of particle detectors that couple to the
Klein-Gordon field\cite{Bir-Dav}. In this interpretation, different choices
for $A$ and $B$ lead to different many-particle theories.  A particularly
simple choice is $A=\half$, $B=-i/2$, in which case the ground state $\xi_0$
is the adiabatic vacuum\cite{Bir-Dav}. Note that the adiabatic vacuum is
among the one-parameter family of vacua that are symmetric about $x_2=0$, in
the sense that
\begin{equation}
\xi_0\left(x_1, x_2\right) =
\overline{\xi_0\left(x_1, -x_2\right)}
\ \ .
\end{equation}

\section{Cosmology and the no-boundary proposal}
\label{sec:cosmology}

As mentioned in the introduction, there exist spatially homogeneous
cosmologies whose dynamics is, modulo certain global caveats, given by our
action. Two such cosmologies are the positive curvature Friedmann model
with a conformally or minimally coupled massless scalar
field\cite{pagescalar}; a third one is the vacuum Kantowski-Sachs
model\cite{rayrev}. The quantum theories constructed in the previous sections
can therefore be interpreted as quantizations of these cosmological models.

This cosmological interpretation raises new questions about the quantum
theories. One such question is whether proposals that have been put forward
for a ``quantum state of the
Universe"\cite{vatican,HH,hawNB,linde,vilePI,vileout,leaking} give quantum
states that belong to any of our Hilbert spaces. In this section we wish to
investigate this question for the no-boundary proposal of Hartle and
Hawking\cite{vatican,HH,hawNB}. For concreteness, we shall first focus on the
positive curvature Friedmann model with a conformally coupled scalar field.
The other two models will be discussed at the end of the section.

Recall that the no-boundary proposal of Hartle and
Hawking\cite{vatican,HH,hawNB} is a topological statement about the manifolds
that are taken to contribute to the path integral in terms of which the wave
function is defined. In the metric variables of general relativity one
writes\cite{hartle_LH,hallhartle/contour,jjhjl3}
\begin{equation}
\Psi_{\rm NB}\left( h_{ij},\phi; \Sigma \right) =
\sum_M \int
{\cal D} g_{\mu\nu} \, {\cal D} \Phi
\exp \left[ - I \left( g_{\mu\nu}, \Phi; M \right) \right]
\ \ ,
\label{nb}
\end{equation}
where $I \left( g_{\mu\nu}, \Phi; M \right)$ is the (Euclidean) action of
the gravitational field $g_{\mu\nu}$ and the matter fields $\Phi$ on the
four-manifold~$M$. The four-manifolds $M$ are required to be compact with a
boundary, such that their boundary is the three-surface $\Sigma$ which
appears in the argument of the wave function. The path integral is over
metrics $g_{\mu\nu}$ and matter fields $\Phi$ on $M$ which induce the values
$h_{ij}$ and $\phi$ on~$\Sigma$. To give a meaning to this formal expression,
additional input is needed. This would include specifying the four-manifolds
$M$ that contribute to the sum and their relative weights, and giving for each
$M$ a definition of the path integral. For further discussion of these general
issues, see for example Ref.\cite{hallhartle/contour}.

We now consider the positive curvature Friedmann model, whose (Lorentzian)
metric is given by
\begin{equation}
ds^2 = {2G\over3 \pi}
\left[ - {\tilde N}^2 (t) dt^2 + a^2 (t) d \Omega_3^2 \right]
\ \ ,
\label{fried_metric}
\end{equation}
where $d\Omega_3^2$ is the metric on the unit three-sphere and $G$ is
Newton's constant. The matter is taken to be a massless, conformally coupled
scalar field~$\Phi$, homogeneous on the constant $t$ surfaces. After the
field redefinition
\begin{equation}
\begin{array}{rcl}
a &=& x_2
\\
\Phi &=&
{\displaystyle{
{\left( 3 \over 4\pi G \right)}^{\!\half} \,
{x_1 \over x_2} }}
\\
\noalign{\smallskip}
{\tilde N} &=& x_2 N
\ \ ,
\end{array}
\label{fried1_redef}
\end{equation}
the action of this model is given by Eq.~(\ref{action}), with
$\delta=0$\cite{HH}.

A delicate question in this field redefinition is the allowed range of the
variables. The usual Hamiltonian formulation of general relativity in the
metric variables assumes that the metric on the three-surfaces is
positive definite. In the Friedmann model this means that the scale factor
should be nonvanishing, and by convention one can take $a>0$. Under
this restriction the field redefinition (\ref{fried1_redef}) is nonsingular,
and the equivalent restriction is $x_2>0$. In the previous sections, however,
it was assumed that the coordinates $x_I$ take all real values. We shall now
proceed under the pretense that these global issues can be ignored and that
the analysis of the previous sections is directly applicable to the
cosmological situation. The validity of this pretense will be addressed at
the end of the section.

Let us first consider the no-boundary proposal in a position
representation. Now $\Sigma=S^3$, and  we take the four-manifold in
Eq.~(\ref{nb}) to be the closed four-dimensional ball~${\bar{B}}^4$\cite{HH}.
The classical solution with the no-boundary data is well known. It is a real
Euclidean solution, and in the Lorentzian conventions used above it can be
written as
\begin{equation}
\begin{array}{rcl}
x_I (t) &=& x_I^b \, e^t
\\
N &=& -i
\ \ ,
\end{array}
\label{pos_nb_sol}
\end{equation}
where $x_I^b$ are constants. From Eqs.\ (\ref{fried_metric}) and
(\ref{fried1_redef}) it is seen that this is just a flat Euclidean metric with
a constant $\Phi$, expressed in a hyperspherical coordinate system where
$e^t$ is the radial coordinate. $t=-\infty$ corresponds to the coordinate
singularity at the center of the~${\bar{B}}^4$, and the boundary $S^3$ is at
$t=0$ with the induced values $x_I^b$ for~$x_I$. The action of this solution
in terms of the boundary data is, in our Lorentzian conventions,
\begin{equation}
S_{class} \left( x_I^b \right)
=
{i\over2} \left( {\left(x_1^b\right)}^2 - {\left(x_2^b\right)}^2 \right)
\ \ .
\end{equation}
Assuming that the no-boundary wave function is well approximated by the
exponential of the negative of the classical Euclidean action, one obtains
the estimate\cite{HH}
\begin{equation}
\Psi_{\rm NB} \left(x_1, x_2 \right)
\sim P \exp \left( -\casehalf x_1^2 + \casehalf x_2^2 \right)
\ \ ,
\label{nbposition}
\end{equation}
where we have dropped the superscript~$b$. The prefactor $P$ is assumed to be
slowly varying compared with the exponential factor.

Two comments here are in order. Firstly, it appears not to
be known whether in this model there exists a set of no-boundary ``initial"
conditions for a minisuperspace path integral such that these conditions
could be stated in terms of acceptable combinations of the initial values of
$x_I$ and~$p_I$. A general discussion of this kind of conditions can be
found in Refs.\cite{jjhjl3,loukoPLB} (see also Ref.\cite{initial}). We shall
therefore not attempt to define a no-boundary wave function in the position
representation beyond the above semiclassical estimate.

Secondly, we have here followed Ref.\cite{HH} in choosing the conventional
sign in the ``Wick rotation" which relates the Lorentzian and Euclidean
theories. This amounts to the relation $iS=-I$ between the Lorentzian and
Euclidean actions. For a discussion of this point, see
Refs.\cite{hallhartle/contour,jjhjl3}.

Return now to those $\delta=0$ quantum theories of section~\ref{sec:position}
where we saw that all states in the Hilbert space are representable by
functions of~$x_I$. Do any of these Hilbert spaces admit wave functions of
the form~(\ref{nbposition})?  The answer is clearly affirmative, as is seen
from the asymptotic properties of the parabolic cylinder
functions\cite{Ab-Steg}: all the Hilbert spaces built on $\vonepos$
(\ref{vonepos}) satisfying the condition (\ref{ABcondition}) contain such wave
functions. Further, the wave function that arguably most closely
conforms to Eq.~(\ref{nbposition}) is the ground state $\xi_0$ annihilated
by~$\jhm$. We have thus found Hilbert spaces in which the ground state
exhibits the asymptotic behavior associated with the semiclassical
no-boundary wave function.

Note that although the leading order exponential term at $x_2\to\infty$ in
these ground state wave functions is real, the wave functions are nevertheless
genuinely complex-valued. Recovering such complex-valued wave functions from
the no-boundary integral (\ref{nb}) by suitable complex integration contours
has been discussed for example in
Refs.\cite{hallhartle/contour,jjhjl3,hartle/contour,jjhjl12}.

None of the Hilbert spaces built on the counterparts of $\vonepos$
with $x_1$ and $x_2$ interchanged admit wave functions with the
semiclassical form~(\ref{nbposition}). The same is true for the Hilbert space
built on ${V^{(1/\sqrt\pi),0}_{\rm pos}}$. Note, however, that  states in the
Hilbert space built on ${V^{(1/\sqrt\pi),0}_{\rm pos}}$ have been suggested
as representing wormhole wave functions\cite{hawpagehole}.

We now turn to the holomorphic representations. Investigating the no-boundary
proposal in these representations is very similar to the analogous problem in
systems involving fermions\cite{death/hall,death/hugh1,death/hugh2,espobook}.

We anticipate that the no-boundary wave function can be written as the
integral
\begin{equation}
\Psi_{\rm NB} = \int
\cD \left( \zi , \zbi , N \right)
\exp\left (i S \right)
\ \ ,
\label{holo_nb}
\end{equation}
where the arguments of $\Psi_{\rm NB}$ are the appropriate ones for the chosen
representation, and $S$ consists of the integral term in the actions
(\ref{holo_action}) and (\ref{mix_action}) with some appropriate boundary
terms. As in the metric variables, we wish to define the integral to be over
fields living on the four-manifold~${\bar{B}}^4$. For simplicity, we continue
to employ a Lorentzian notation. Whether the configurations to be integrated
over are Lorentzian or Euclidean or complex would depend on the precise
definition to be given to the integral.

We would like to define the expression (\ref{holo_nb}) to be compatible
with the classical system in the sense that the classical solutions to the
corresponding boundary value problem be well-defined. In a Berezin-type path
integral for fermionic systems, such a condition has been applied to the
no-boundary proposal in
Refs.\cite{death/hall,death/hugh1,death/hugh2,espobook}. We shall adopt the
analogous condition here. To make this precise, recall that
Eqs.~(\ref{zeds}) can be inverted to give
\begin{equation}
\begin{array}{rcl}
x_I &=& {\displaystyle{
{1\over \sqrt{2}} \left( \zi + \zbi \right) }}
\\
\noalign{\smallskip}
p_I &=&
{\displaystyle{
{i\over \sqrt{2}} \left( \zi - \zbi \right) }}
\ \ .
\end{array}
\label{inv_zeds}
\end{equation}
We therefore define a solution to the holomorphic equations of motion to be
regular if the corresponding four-dimensional gravity and matter fields
computed from Eqs.\ (\ref{fried1_redef}) and (\ref{inv_zeds}) are regular
on~${\bar{B}}^4$.

Consider now the general solution to the holomorphic equations of motion,
\begin{equation}
\begin{array}{rcl}
z_1 &=& C_1 \, e^{i\tau}
\\
{\bar{z}}_1 &=& C_2 \, e^{-i\tau}
\\
z_2 &=& C_3 \, e^{-i\tau}
\\
{\bar{z}}_2 &=& C_4 \, e^{i\tau}
\ \ ,
\end{array}
\label{zsolution}
\end{equation}
where $C_i$ are constants, $\tau=\int^t N(t') dt'$, and $N$ is allowed to
take general complex values. With the choice $N=-i$, a regular classical
solution is obtained by setting $C_2=C_3=0$. This is the holomorphic
counterpart of the solution~(\ref{pos_nb_sol}); note that both the metric
and the matter field in this solution are complex-valued for generic values
of $C_1$ and~$C_4$. $t=0$ gives again the boundary of~${\bar{B}}^4$, and
$t=-\infty$ gives the coordinate singularity at the center. The corresponding
classical boundary value problem in terms of $\zi$ and $\zbi$ consists of
setting the initial values of ${\bar{z}}_1$ and $z_2$ to zero, and specifying
freely the final values of $z_1$ and~${\bar{z}}_2$. The action appropriate
for this boundary value problem is given by Eq.~(\ref{mix_action}). One is
therefore led to expect that the no-boundary integral is given by the
amplitude $G \! \left( \alpha_1, {\bar\alpha}_2 ; {\bar\beta}_1, \beta_2
\right)$~(\ref{mix_pi}), with ${\bar\beta}_1=0=\beta_2$. In particular, this
means that the no-boundary integral gives a wave function in the
holomorphic-antiholomorphic representation. This is analogous to what happens
in the massless fermionic systems in
Refs.\cite{death/hall,death/hugh1,death/hugh2}.

Let us try to evaluate this no-boundary integral. Consider first a
semiclassical estimate. The classical action (\ref{mix_action}) evaluated on
a solution for which the initial values of ${\bar{z}}_1$ and $z_2$ vanish is
equal to zero, and the semiclassical estimate to the integral therefore
consists entirely of a prefactor. Such an estimate is compatible with many
states in the Hilbert space $\hilonemix$, arguably most closely with the
ground state $\chi_0(z_1,{\bar{z}}_2)=1/\bar{z}_2$ annihilated by~$\jhm$.
For concreteness, we concentrate here on~$\hilonemix$; an analogous
discussion holds for~$\hiltwomix$.

Consider then the exact amplitude evaluated in
subsection~\ref{subsec:path-mix}. The above discussion implies that the
no-boundary state should be given by
\begin{equation}
\Psi_{\rm NB}
\left( z_1, {\bar{z}}_2 \right) =
\lim_{\textstyle{{ {\bar\beta}_1\to0 \atop
\beta_2\to0 }}}
\psibetas
\left( z_1, {\bar{z}}_2 \right)
\ \ ,
\label{nb_mix}
\end{equation}
where $\psibetas$ is given by~(\ref{psibetas}). Unfortunately, even though
$\psibetas$ is in $\hilonemix$ for $|{\bar\beta}_1/\beta_2|<1$, the limit
(\ref{nb_mix}) does not yield a vector in~$\hilonemix$, as it is seen from
Eqs.\ (\ref{G:mix}) and (\ref{Psi_product}) that $\left(\psibetas,
\psibetas\right)$ diverges no matter how ${\bar\beta}_1$ and $\beta_2$ are
taken to zero.

Thus, although a semiclassical estimate for the no-boundary integral is
compatible with some states in the Hilbert space~$\hilonemix$, and
arguably most closely with the ground state, the exact path measure of
subsection \ref{subsec:path-mix} does not yield a well-defined no-boundary
wave function. An attempt to improve this situation might be to argue that
before taking the limit in Eq.~(\ref{nb_mix}), it is legitimate to
renormalize $\psibetas$ by a factor which only depends on ${\bar\beta}_1$
and~$\beta_2$. A suitable choice for this renormalization factor and a
suitable prescription for taking ${\bar\beta}_1$ and $\beta_2$ to zero then
produces the ground state wave function, as is easily seen from
Eq.~(\ref{displacement1}). It is, however, not clear whether a prescription
of this kind can be justified from the path integral.

The above discussion was based on the regular classical solution in which
$N=-i$ and $C_2=C_3=0$ in~(\ref{zsolution}). There exists a second regular
solution, obtained with $N=-i$ and $C_1=C_4=0$. In this solution, $t$ ranges
from $t=0$ at the boundary of~${\bar{B}}^4$ to $t=+\infty$ at the coordinate
singularity at the center of~${\bar{B}}^4$. The no-boundary path integral
with the boundary data compatible with this solution gives a wave function in
the antiholomorphic-holomorphic representation. It was seen in
section~\ref{sec:mix} that the Hilbert spaces obtained in the
antiholomorphic-holomorphic representation with $\delta=0$ are isomorphic to
those obtained in the holomorphic-antiholomorphic representation. This
implies that a discussion of the antiholomorphic-holomorphic no-boundary wave
function proceeds exactly as in the holomorphic-antiholomorphic case above.

Note that once the relation between the Lorentzian and Euclidean theories
has been fixed to be the conventional one, the choice between a classical
solution defined for $-\infty<t\le0$ or $0\le{t}<\infty$ does not make a
difference in the semiclassical estimate for the no-boundary wave function in
the metric variables. In the holomorphic variables, however, this choice picks
respectively the holomorphic-antiholomorphic or antiholomorphic-holomorphic
representation. A similar observation was made in a supersymmetric system in
Ref.\cite{death/hugh2}.

Return now to the restriction $x_2>0$, which has been ignored so far. In the
Dirac quantization, most of the analysis of the previous sections appears to
go through unperturbed when this restriction is introduced, both in the
holomorphic representations and the position representation. Notice in
particular that the allowed values of the $J$'s (\ref{Jclass}) are
unaffected. The only step where the restriction seems to make an essential
difference is when one attempts to define integral formulas for the inner
product. The formula (\ref{pos_ip2}) for the Klein-Gordon inner product in
the position representation remains nevertheless valid. The restriction
$x_2>0$ appears also to be consistent with all the steps in finding the
semiclassical estimates to the no-boundary wave functions, both in the
position representation and the holomorphic representations.

On the other hand, the step that seems unlikely to remain justified under
the restriction $x_2>0$ is the construction of the exact holomorphic path
measures in subsection~\ref{subsec:can-mix}. Our method relied on introducing
first an unconstrained path integral for each oscillator, and finally
integrating over the Lagrange multiplier to enforce the constraint. The
restriction $x_2>0$ would affect the unconstrained path integral for the
second oscillator, and hence most likely the final result.

To end this section, let us briefly discuss two other cosmological models
whose dynamics is described by our action. First, consider again the positive
curvature Friedmann model~(\ref{fried_metric}), but now with a massless,
minimally  coupled scalar field~$\phi$, homogeneous on the constant $t$
surfaces. The field redefinition mapping this system to our model with
$\delta=0$ is\cite{pagescalar}
\begin{equation}
\begin{array}{rcl}
a &=& \sqrt{x_2^2 - x_1^2}
\\
\noalign{\smallskip}
\phi &=& {\displaystyle{
{\left( 3 \over 4\pi G \right)}^{\!\half} \,
\mathop{\rm artanh}\nolimits
\left({x_1 \over
x_2}\right) }}
\\
\noalign{\medskip}
{\tilde N} &=& \sqrt{x_2^2 - x_1^2} \, N
\ \ ,
\end{array}
\end{equation}
and the restriction for the coordinates is $x_2>|x_1|$. The Dirac
quantization proceeds as above; however, the formula (\ref{pos_ip2}) for the
Klein-Gordon inner product is now not compatible with the restriction.
Similarly, the semiclassical estimates to the no-boundary wave function with
the four-manifold ${\bar{B}}^4$ are exactly as above, implying
compatibility of the Hilbert spaces and the semiclassical no-boundary wave
function. An attempt to find an exact holomorphic no-boundary wave function
using the measure of subsection \ref{subsec:path-mix} faces the same
difficulty as with the conformally coupled scalar field.

Finally consider the vacuum Kantowski-Sachs model. The metric is given by
\begin{equation}
ds^2 = {G\over2 \pi}
\left[ - {\tilde N}^2 (t) dt^2 + a^2 (t) d\chi^2 + b^2 (t) \Omega_2^2 \right]
\ \ ,
\end{equation}
where $d\Omega_2^2$ is the metric on the unit two-sphere and $\chi$ is an
angular coordinate with period~$2\pi$. The field redefinition leading to our
action with $\delta=0$ is\cite{rayrev}
\begin{equation}
\begin{array}{rcl}
a &=&
{\displaystyle{
{x_2 - x_1  \over x_2 + x_1 } }}
\\
\noalign{\smallskip}
b &=& \half {(x_2 + x_1)}^2
\\
\noalign{\medskip}
{\tilde N} &=&
{(x_2 + x_1)}^2 N
\ \ ,
\end{array}
\end{equation}
and the restriction is again $x_2>|x_1|$. The Dirac quantization proceeds
thus as in the Friedmann model with the minimally coupled scalar
field. Now, however, there are two relevant four-manifolds that admit
no-boundary type classical solutions\cite{york,raythesis}. For the
four-manifold  ${\bar{B}}^3\times{}S^1$, the analysis proceeds as for
the Friedmann model with a minimally coupled field. For the four-manifold
${\bar{B}}^2\times{}S^2$, on the other hand, the semiclassical estimate to
the no-boundary wave function in the metric variables is more
complicated\cite{jjhjl3,york,raythesis,loukoAP}, and it is not
obvious how to formulate the corresponding variational problem in the
holomorphic variables. We shall not attempt to discuss this manifold further.

\section{Conclusions and discussion}
\label{sec:conclusions}

In this paper we have discussed the Dirac quantization and path integral
quantization of a simple quantum mechanical system with a quadratic
Hamiltonian constraint. The system consists of two harmonic oscillators with
identical frequencies, and the dynamics is given by a Hamiltonian
constraint which sets the energy difference of the oscillators equal to a
prescribed real number. In the Dirac quantization we employed a class of
representations which generalize Bargmann's coherent state representation for
an ordinary unconstrained oscillator. In each of our representations an inner
product was uniquely determined, under certain assumptions, by imposing
Hermitian conjugacy relations on a suitable operator algebra. The sets of the
Hilbert spaces obtained in the different representations were found to have
drastically different properties; in particular, some of these Hilbert spaces
contradicted the classical expectations for the spectra of certain operators.
The analogous Dirac quantization in a conventional position representation
was shown to give Hilbert spaces that are isomorphic to those obtained in the
coherent state representations in which the spectra of the operators do not
violate classical expectations. Carefully defined coherent state path
integrals were shown to yield the matrix elements of the identity operator in
the Hilbert spaces obtained in Dirac's quantization, with respect to an
overcomplete basis of representation-dependent generalized coherent states.
Finally, the results were interpreted as quantizations of a class of
cosmological models. Both in the position representation and in the relevant
coherent state representations, the ``ground state" in the appropriate
Hilbert spaces was shown to have the semiclassical behavior expected of the
no-boundary wave function of Hartle and Hawking.

Our Dirac quantization was carried out following closely the algebraic
quantization program of Ashtekar\cite{AAbook}. This program was previously
applied to our model in Refs.\cite{AAbook,tate:th}, and we took our classical
Poisson bracket algebras and quantum commutator algebras to be directly the
same as in that earlier work. The point where we differed from the earlier
work was the choice of the vector spaces on which these algebras, most
importantly the physical operator algebra~$\astarphy$, are represented.
Whereas $\astarphy$ in Refs.\cite{AAbook,tate:th} was represented on a vector
space spanned by generalized angular momentum eigenstates, we took the vector
space to consist of certain functions of two real or complex variables. After
finding the inner product, the Cauchy completion was seen to lead precisely
to the set of abstract Hilbert spaces obtained in Refs.\cite{AAbook,tate:th};
however, some of these spaces contained vectors that are no longer
representable by functions. We then restricted the attention to the Hilbert
spaces where all the vectors are representable by functions, and found that
this set contains the most physically interesting spaces. In particular, this
set contains the spaces where the spectrum of the operator
$\jho$~(\ref{Jquantumo}) is bounded below, in agreement with the lower bound
for the corresponding classical function $J_0$~(\ref{Jclasso}).

Our representations on functions of complex variables generalized
Bargmann's coherent state representation for a single unconstrained
oscillator in that we allowed the functions to be either analytic or
antianalytic in each of the arguments. Analyticity in both arguments led
precisely to those spaces where the spectrum of $\jho$ satisfies the
classical expectations, whereas antianalyticity in both arguments led only
to spaces where these classical expectations are violated. Among the
mixed representations, on the other hand, the spaces where the classical
expectations for $\jho$ are satisfied are distributed so that the
eigenenergies of the analytically represented oscillator are always the
positive half-integers.

Although the representations thus exhibit a clear disparity between
analyticity and antianalyticity, it is interesting that this disparity is not
as severe as in the case of an unconstrained harmonic oscillator. For the
unconstrained oscillator, the standard Bargmann theory is recovered by
considering the operator algebra generated by the operators $\left\{{\hat z},
\hat{\bar z}, {\hat{\openone}}\right\}$ with commutation relations analogous
to Eqs.~(\ref{zedcomms}), with a representation analogous to that in
Eqs.~(\ref{zedrep1}) on the vector space  $\hbox{Span} \left\{z^k \mid
k=0,1,2,\ldots \right\}$. The standard inner product is uniquely recovered by
imposing the Hermitian conjugacy relation ${\hat z}^\dagger = \hat{\bar z}$.
If one instead tries to represent this algebra antianalytically, in analogy
with Eqs.~(\ref{zedrep2}) on a vector space spanned by powers of~$\bar z$,
there will be no inner product compatible with the Hermitian conjugacy
relation. An inner product can however be found if one adopts the
antianalytic representation on a vector space involving not functions but
(anti)holomorphic distributions\cite{holodist}.

In the case where the energy difference $2\delta$ is an integer, it is clear
how to build a quantum theory for our system starting from the conventional
quantum theory of two unconstrained oscillators. Our method recovered
quantum theories for arbitrary real values for~$2\delta$, and for integer
values of~$2\delta$ one of our theories was indeed isomorphic to the expected
theory. It is natural to ask whether the theories with integer values
of~$2\delta$ are distinguished by some special properties. One way to look
at this is to recall that our Hilbert spaces carry representations
of the Lie algebra of $SO(2,1)$. By exponentiation this means that the spaces
carry representations of the universal covering group of $SO_c(2,1)$,
but not necessarily those of $SO_c(2,1)$ or its double cover $SU(1,1)$.
(Here $SO_c(2,1)$ stands for the connected component of $SO(2,1)$.)
However, our spaces with integer eigenvalues of $\jho$ do carry
unitary representations of $SO_c(2,1)$, and those with half-integer
eigenvalues of $\jho$ carry unitary representations of
$SU(1,1)$\cite{bargmann-lorentz}. Within the spaces in which
the spectrum of $\jho$ is bounded below, this means that representations of
$SO_c(2,1)$ are obtained in $\hilpholo$ with $2\delta$ odd, and representations
of $SU(1,1)$ are obtained in $\hilpholo$ with $2\delta$ even.

Another way to look at this question is to consider a modified theory where we
add to the system a true classical Hamiltonian equal to~$J_0$. The constraint
remains first class\cite{dirac}, and from Eqs.~(\ref{Cso21alg}) it is seen
that the classical motion is periodic in $t$ with period~$2\pi$. Taking the
quantum Hamiltonian to be $\jho$~(\ref{Jquantumo}), the Hilbert spaces
are the same as before, but now there is a Schr\"odinger equation giving these
spaces nontrivial dynamics. For concreteness, let us only consider the
spaces~$\hilpmholo$, in which the spectrum of $\jho$ is bounded below. The
Schr\"odinger evolution in $\hilpmholo$ is not periodic for general values
of~$\delta$; however, it is periodic with period~$2\pi$ up to the phase
$e^{i\pi(-1\mp2|\delta|)}$. Integer values of $2\delta$ (which by definition
of $\hilpmholo$ only can occur with the upper sign) are distinguished by
making this phase real. A related discussion involving a parity operator has
been given in Refs.\cite{AAbook,tate:th}.

It is also possible to do the path integrals of subsections
\ref{subsec:path-holo} and \ref{subsec:path-mix} in the presence of $J_0$ as
the true Hamiltonian. The propagation amplitudes analogous to Eqs.\
(\ref{G:holo}) and (\ref{G:mix}) are
\begin{equation} G_\pm \!
\left( \alpha_I ; {\bar \beta}_I ; t \right)
=
e^{-it/2}
{\left(
\alpha_1 {\bar \beta}_1
\over
\alpha_2 {\bar \beta}_2
\right)} ^\delta
I_{\pm2|\delta|}
\left( 2 e^{-it/2} \sqrt{\alpha_1 \alpha_2
{\bar \beta}_1 {\bar \beta}_2 }
\right)
\end{equation}
and
\begin{equation}
G \! \left( \alpha_1, {\bar\alpha}_2 ;
{\bar\beta}_1, \beta_2 ; t \right)
=
{\Gamma(1-2\delta) \over
{\left(
{\bar\alpha}_2 \beta_2 e^{it/2} -\alpha_1 {\bar \beta}_1 e^{-it/2}
\right)}^{1-2\delta}}
\ \ .
\end{equation}
These amplitudes are readily recognized as the matrix elements of the
time evolution operator $\exp\left(-it\jho\right)$. Note that they are
periodic in $t$ with period $2\pi$ up to the $\delta$-dependent phase.

As has been emphasized in Ref.\cite{tate:th}, it is remarkable that there
emerged no ``quantization condition" for the allowable values of~$\delta$.
One is inclined to relate this to the fact that the reduced phase space of
the classical system has infinite volume, as was seen at the end of
section~\ref{sec:model}. When a similar analysis is carried out in a model
where the energy difference constraint is replaced by an energy sum
constraint, so that in particular the reduced phase space has finite volume,
one indeed finds that there is a quantization condition for the allowed
values of the energy sum\cite{tate:th,stillerman,rovelli1,unruh}.  The
relation of the different representations in the energy sum model  is very
similar to that in the energy difference model, and also coherent state path
integrals can be constructed very much as in the energy difference model. We
shall briefly outline these results in the appendix.

On the other hand, both the energy difference model and the energy sum model
are rather special among all quadratic super-Hamiltonians in that the
frequencies of the two oscillators were fine-tuned to be equal. It would be
of interest to understand whether the approach in this paper could be adapted
to the more general case where the frequencies are different. One might
anticipate problems at least when the ratio of the the two frequencies
is irrational, since the classical motion in that case is highly
chaotic\cite{haji}.

In sections \ref{sec:holo} and~\ref{sec:mix}, we defined the formal
holomorphic path integral expressions so as to obtain quantities that are
related to the Hilbert spaces. The definition involved several delicate
points, perhaps most crucially the assumption that the integral over the
Lagrange multiplier can be pulled outside the  $\cD \left( \zi , \zbi\right)$
integrals, and that the contour for the Lagrange multiplier can then be
chosen complex. A similar method has been used in quantum cosmology, and the
possible caveats of this method discussed in that context in
Refs.\cite{jjhjl3,initial} may well be relevant also here.

It might be of interest to investigate in more detail also conventional
configuration space path integrals in our system, and their relation to the
Hilbert spaces obtained in the position representation in
section~\ref{sec:position}. Some issues involving the definition of such
integrals in this model have been discussed in Ref.\cite{LafLouko}.

In section \ref{sec:cosmology} we saw that our model can be reinterpreted
as a quantization of certain cosmological models, provided the range of the
configuration space variables is suitably restricted. Whilst such a
restriction seemed to have little effect on the Dirac quantization, one
nevertheless expects that the operators of physical interest in the quantized
cosmologies should be constructed in a way which takes into account the
ranges of the variables. One suggestion to construct such operators is via
the ``evolving constants of
motion"\cite{tate:th,rovelli1,rovelli2,carlip,atu} (see, however, the
discussion in Ref.\cite{kuchar_winn}).

The cosmological restriction on the range of the configuration space
variables was not compatible with the construction of the exact
holomorphic path measures in subsection~\ref{subsec:can-mix}. In view of
this, one is tempted to regard our affirmative results about the
compatibility of the Hilbert spaces and the semiclassical no-boundary wave
function as more interesting than the failure of the measure of
subsection~\ref{subsec:can-mix} to yield a well-defined exact no-boundary
wave function. In the Friedmann model with a conformally coupled scalar
field, it might be possible to investigate this issue further in a
representation where the wave functions are analytic on a
half-plane\cite{paul}.

In the cosmological interpretation, our approach bears some similarity to
the quantization of spatially homogeneous cosmologies in the connection
representation of Ashtekar's variables\cite{AAbook,ash_pull,kodama}.
There are, however, two significant differences.

Firstly, recall that the functions $\zi$ and $\zbi$ on the classical phase
space take arbitrary complex values, or under the cosmological
restrictions arbitrary values in certain open sets in the complex plane. This
makes it natural to look for a quantum representation where the wave functions
are analytic or antianalytic functions of complex variables. On the other
hand, when spatially homogeneous cosmologies are written in terms of
Ashtekar's variables, the real part of the Ashtekar connection $A$ is a
constant function over the classical Lorentzian phase
space\cite{ash_pull,kodama,bom_tor}. The value of this constant is determined
by the homogeneity type; for example in Bianchi type I the real part of $A$
simply vanishes. One would therefore expect the $A$-representation of the
quantum theory to be more analogous to a conventional momentum representation
than to a coherent state representation, and this makes it unclear whether
the wave functions should be analytic in~$A$. A specific proposal for
analyticity was made in Ref.\cite{kodama}.

Secondly, recall that for our semiclassical estimate to the no-boundary
wave function we had to solve a classical boundary value problem. In the
mixed representations, it was found that the relevant boundary value problem
had a solution for arbitrary complex values of the boundary data compatible
with the restriction for the range of the variables. In terms of Ashtekar's
variables, however, the corresponding classical boundary value problem is more
problematic. Consider for concreteness the Euclidean version of the theory,
where $A$ classically takes generic real values, and suppose that the
cosmological constant vanishes. In the cosmological models that have been
investigated, the no-boundary classical solutions are such that
four-dimensional regularity at the ``center" of the four-manifold fixes the
components of the ``initial" connection to certain numerical
values\cite{loukoPLB}. As Ashtekar's super-Hamiltonian consists of a pure
kinetic term with a Lorentz signature supermetric, this means that solutions
to the corresponding boundary value problem  with a specified boundary
connection exist only when this boundary connection lies on the null cone
(with respect to the supermetric) of the ``initial" connection. Thus, for a
given generic boundary connection, solutions to the boundary value problem do
not exist. It is unclear what this should be taken to imply for the
semiclassical estimate to a no-boundary wave function in the connection
representation of the Ashtekar variables.

One should emphasize that these two points rely heavily on the restriction
of the Ashtekar variables to spatially homogeneous cosmologies. It is not
clear whether these points remain relevant when discussing the connection
representation without such symmetry assumptions.

Finally, we would like to briefly discuss the relation of semiclassical
methods to our Dirac quantization. In much of the work on quantum
cosmology, physically interesting solutions to the quantum Hamiltonian
constraint have been sought in the form of a Born-Oppenheimer
approximation, the celebrated result being the derivation of the
functional Schr\"odinger equation of quantum field theory on curved
spacetime from the Wheeler-DeWitt equation\cite{qftics}. This method has
been criticized\cite{kuchar_winn}, among other things on the grounds that
one does not assume the original wave function to lie in a Hilbert space,
but rather one only later introduces additional approximate normalization
conditions for parts of the wave function to recover physical predictions.
It therefore remains unclear what the theory is that the approximation
seeks to approximate. This leads to problems when discussing for example
superpositions of solutions.

Suppose now that one wishes to use the Born-Oppenheimer method in our
model. Does one recover physical predictions that agree with those in any of
our exact Hilbert spaces?

Mimicking the usual approach in the cosmological case, let us consider
quantum theory in the position representation. We wish to assume that the
second oscillator dominates the dynamics in some appropriate sense, and that
the first oscillator can be treated as a small perturbation. We therefore
expect the energy of the second oscillator to be large compared with that of
the first one, so that $\delta$ is large and negative. To implement this, we
write $\delta=-{1\over4}M\gamma$, where $\gamma$ is considered a fixed
positive number, and the limit of interest will be that of large
positive~$M$. We also perform the rescaling  $x_2\to{M}^{1/2}x_2$,
$p_2\to{M}^{-1/2}p_2$; this is just a canonical transformation and in no way
changes the physics. The classical constraint is then given by
\begin{equation}
\cH =  \half \left[
- {p_2^2 \over M} + M \left( \gamma - x_2^2 \right)
\right]
+
\casehalf \!
\left( p_1^2 + x_1^2\right)
\ \ .
\end{equation}
The quantum constraint obtained by the substitution $p_I^2
\to -\partial^2/\left(\partial x_I^2\right)$ is of a form in which the
Born-Oppenheimer approximation can be implemented by the standard formal
expansion at $M\to\infty$\cite{kuchar_winn,qftics}. In the next-to-leading
order in this expansion one recovers the ordinary Schr\"odinger equation for
the first oscillator, and if one requires the solutions to this equation to be
normalizable in the usual Schr\"odinger inner product, one obtains the
prediction that the energy eigenvalues of the first oscillator are just the
positive half-integers\cite{singh}. This prediction is in exact agreement
with that obtained in our Hilbert space~$\hilonemix$.

On the other hand, it has been suggested that in the quantum cosmological
context one could derive quantum gravitational corrections to quantum
field theory in curved spacetime by carrying out the Born-Oppenheimer
expansion beyond the next-to-leading order\cite{singh,kiefer-singh}.
When applied to our model, this method has been argued\cite{singh} to
be consistent with a higher order correction in $1/M$ to the
half-integer eigenenergies of the first oscillator. Such a correction would
clearly no longer be in agreement with the exact Hilbert
space~$\hilonemix$. (Note, however, that the argument in the form given in
Ref.\cite{singh} relies on introducing in Eq.~(20) therein a quantity
$F(q)$ which is formally divergent.)

Thus, the Born-Oppenheimer approximation gives, at least in the
next-to-leading order, predictions that are consistent with those obtained in
one of our exact Hilbert spaces. One may see this as encouraging for the
Born-Oppenheimer approximation, and one might wish to use the position
representation realizations of this Hilbert space that were constructed in
section \ref{sec:position} to further examine the range of validity of the
approximation. An open question in such an approach would however be how to
distinguish effects due to the the simplicity of our model from potentially
more general effects.

\acknowledgments

I would like to thank Abhay Ashtekar, Claus Kiefer, Don Marolf, Jonathan
Simon, Tejinder Singh and Chrysis Soteriou for discussions on various parts
of this work. I would especially like to thank Ranjeet Tate for numerous
discussions and for making a preliminary version of Ref.\cite{tate:th}
available. This work was supported in part by the NSF grant PHY90-16733 and
by research funds provided by Syracuse University.

\appendix
\section*{Energy sum model}

The methods used in this paper can be applied without essential changes in a
model in which the energy difference constraint (\ref{cH}) is replaced by the
energy sum constraint\cite{tate:th,stillerman,rovelli1,unruh}
\begin{equation}
\tilde{\cal H}
= \casehalf
\left( p_1^2 + p_2^2  + x_1^2 + x_2^2 \right)
-E
\ \ ,
\label{tcH}
\end{equation}
where $E$ is an arbitrary real number. In this appendix we shall outline the
results.

The physical operator algebra $\tastarphy$ is now generated by the set
$\left\{\lhpm, \lho, {\hat{\openone}} \right\}$, where
\begin{equation}
\begin{array}{rcl}
\lhp &=& {\hat z}_1 {\hat {\bar z}}_2\\
\lhm &=& {\hat z}_2 {\hat {\bar z}}_1\\
\lho &=& \half \left(
{\hat z}_1 {\hat{\bar z}}_1 -
{\hat z}_2 {\hat{\bar z}}_2  \right)
\ \ ,
\end{array}
\end{equation}
and the commutators are
\begin{equation}
\begin{array}{rcl}
\left[\lhp,\lhm\right] &=& 2 \lho
\\
\noalign{\smallskip}
\left[\lho,\lhpm\right] &=& \pm \lhpm
\ \ .
\end{array}
\end{equation}
The $\hat L$'s are recognized as the generators of the Lie algebra of
$SU(2)$. With the factor ordering
\begin{equation}
\hat{\tilde{\cal H}} =
{\hat z}_1 {\hat{\bar z}}_1
+{\hat z}_2 {\hat{\bar z}}_2
+ (1-E) {\hat{\openone}}
\ \ ,
\end{equation}
the constraint $\hat{\tilde{\cal H}}\psi=0$ is easily solved in the
doubly holomorphic, doubly antiholomorphic, and mixed representations, and
the irreducible representations of $\tastarphy$ on vector spaces analogous
to those in the energy sum model are easily found. The Hermitian conjugacy
relations inherited from the complex conjugation properties of the classical
counterparts of the $\hat L$'s are
\begin{equation}
{\hat L}_{\pm}^{\dagger} = {\hat L}_{\mp} \ ,
\qquad
{\hat L}_0^{\dagger} = {\hat L}_0
\ \ .
\label{Ldag}
\end{equation}

In the doubly holomorphic representation, one obtains quantum theories only
for positive integer values of~$E$, and the Hilbert spaces are just the usual
finite dimensional spaces carrying unitary representations of the
group~$SU(2)$. This reproduces the result of
Refs.\cite{tate:th,stillerman,rovelli1,unruh}. Taking $E$ to be a positive
integer, the path integral analogous to that in subsection
\ref{subsec:can-holo} yields the amplitude
\begin{equation}
G \!
\left( \alpha_I ; {\bar \beta}_I \right)
=
{1\over (E-1)!}
{\left(
\alpha_1 {\bar \beta}_1
+
\alpha_2 {\bar \beta}_2
\right)} ^{E-1}
\ \ ,
\end{equation}
where the integration contour has been chosen as for integer values of
$2\delta$ in subsection~\ref{subsec:can-holo}. This amplitude gives
the matrix elements of the identity operator in the relevant Hilbert spaces
in the overcomplete basis formed by the displacement operator coherent states
for $SU(2)$\cite{klauder-skager,perelomov}.

In the doubly antiholomorphic representation one recovers quantum theories
only for negative integer values of~$E$. A path integral can be constructed as
above. These Hilbert spaces carry unitary representations of $SU(2)$ that are
isomorphic to those in the doubly holomorphic representation; however, the
classical system has no solutions for negative~$E$. The relation between the
doubly holomorphic and doubly antiholomorphic representations is therefore
very similar to that in the energy difference model.

In the mixed representations there exist no inner products that would
satisfy the Hermitian conjugacy relations~(\ref{Ldag}), and no Hilbert
spaces are obtained.

\end{document}